\documentclass[12pt]{article}

 \usepackage{amssymb,amsmath,slashed}
 \usepackage{epsf}
 \usepackage{latexsym}
 \usepackage{epsf}
 \usepackage{amssymb}
 \usepackage{cite}
 \usepackage{epsfig}

\begin{document}

 \rightline{UG-03/07}
 \rightline{KCL-MTH-03-17}
 \rightline{\tt hep-th/0312102}
 \thispagestyle{empty}
 \vskip 1truecm

 \centerline{\Large {\bf Transient Quintessence from Group Manifold Reductions}}
 \vskip .7truecm
 \centerline{\Large\bf or}
 \vskip .7truecm
 \centerline{\Large {\bf how all roads lead to Rome}}

 \vskip 2truecm

 \centerline{\bf E.~Bergshoeff${}^{\dagger}$, A.~Collinucci${}^{\dagger}$, U.~Gran${}^{\ddagger}$,
                 M.~Nielsen${}^{\dagger}$ and D.~Roest${}^{\dagger}$}

 \bigskip

 \centerline{${}^{\dagger}$\ Centre for Theoretical Physics, University of Groningen}
 \centerline{Nijenborgh 4, 9747 AG Groningen, The Netherlands}
 \centerline{E-mail: {\tt (e.a.bergshoeff, a.collinucci, m.nielsen, d.roest)@phys.rug.nl}}
 \vskip .2truecm
 \centerline{${}^{\ddagger}$\ Department of Mathematics, King's College London}
 \centerline{Strand, London WC2R 2LS, United Kingdom }
 \centerline{E-mail: {\tt ugran@mth.kcl.ac.uk}}

 \vskip 2truecm

 \centerline{ABSTRACT}
 \bigskip

We investigate the accelerating phases of cosmologies supported by a metric, scalars and a single
exponential scalar potential. The different solutions can be represented by trajectories on a
sphere and we find that quintessence happens within the ``arctic circle" of the sphere.

Furthermore, we obtain multi-exponential potentials from 3D group manifold reductions of gravity,
implying that such potentials can be embedded in gauged supergravities with an M-theory origin.
We relate the double exponential case to flux compactifications on maximally symmetric
spaces and S-branes. In the triple exponential case our analysis suggests the existence of two
exotic S($D-3$)-branes in $D$ dimensions.

\vfill\eject

\section{Introduction}\label{intro}

Since the observational evidence for an accelerating universe and
a nonzero cosmological constant
\cite{Riess:1998cb,Perlmutter:1998np}, there has been a growing
interest in finding (stable or unstable) De Sitter solutions
\cite{Kallosh:2001gr,Fre:2002pd,deRoo:2002jf,Kachru:2003aw,deRoo:2003rm}
or more general accelerating cosmologies from M-theory
\cite{Townsend:2001ea,Kallosh:2002gg,Ohta:2003uw,Townsend:2003fx,Ohta:2003pu,
Roy:2003nd,Wohlfarth:2003ni,Emparan:2003gg,Ohta:2003ie,Chen:2003ij,Gutperle:2003kc,Chen:2003dc,Wohlfarth:2003kw,
Townsend:2003qv,Neupane:2003pw,Jarv:2003qy,Neupane:2003cs,Vieira:2003tx} (for
general results on accelerating cosmologies, i.e.~earlier and/or
without an M-theory origin, see
\cite{Demaret:1985js,Halliwell:1987ja,Burd:1988ss,Aguirregabiria:1993pm,Lu:1997jk,
Poppe:1997uz,Lu:1996er,Coley:1997nk,Liddle:1998jc,Malik:1998gy,Copeland:1999cs,Green:1999vv,Billyard:1999dg,Coley:1999mj, 
Aguirregabiria:2000hx, Heard:2002dr,Guo:2003rs}).

One of the purposes of this paper is to investigate the
possibility of transient acceleration, i.e.~a period of
quintessence\footnote{Quintessence corresponds to the universe
being filled with a perfect fluid with equation of state
$p=\kappa\,\rho$, where $-1\leq\kappa<-1/3$. This induces
acceleration.}, for a large class of cosmologies whose solutions
are described by a metric and $N$ scalars. We assume that the
scalar potential is given by a single exponential. This has the
consequence that effectively the scalar potential depends on only
one scalar. All other $N-1$ scalars are represented by their
kinetic terms only. Since the metric cannot distinguish between
these different $N-1$ scalars, there is no qualitative difference
between the $N=2$ scalar cosmology and the $N>2$ scalar
cosmologies. We therefore only consider the one-scalar ($N=1$) and
two-scalar ($N=2$) cosmologies.

The cosmological solutions discussed in this paper have been given sometime ago
\cite{Gavrilov:1994sv,Aguirregabiria:2000hx}. The fact that these cosmologies, for particular cases
at least, exhibit a period of quintessence was noted recently in \cite{Townsend:2003fx} where a
specific class of solutions was obtained by compactification over a compact hyperbolic space (for
earlier discussions, see \cite{Lu:1997jk,Poppe:1997uz,Lu:1996er,Aguirregabiria:2000hx,Cornalba:2002fi}). The
relation with S-branes was subsequently noted in \cite{Ohta:2003pu,Roy:2003nd,Ohta:2003ie} (for
general literature on S-brane solutions, see
\cite{Ivashchuk:1989gz,Bleyer:1994xg,Lu:1997jk,Lukas:1997iq,Lu:1996er,
Ivashchuk:1999xp,Gutperle:2002ai,Chen:2002yq,
Kruczenski:2002ap,Deger:2002ie,Wohlfarth:2003ni,Emparan:2003gg,Wohlfarth:2003kw}).

\begin{figure}[h]
\centerline{\epsfig{file=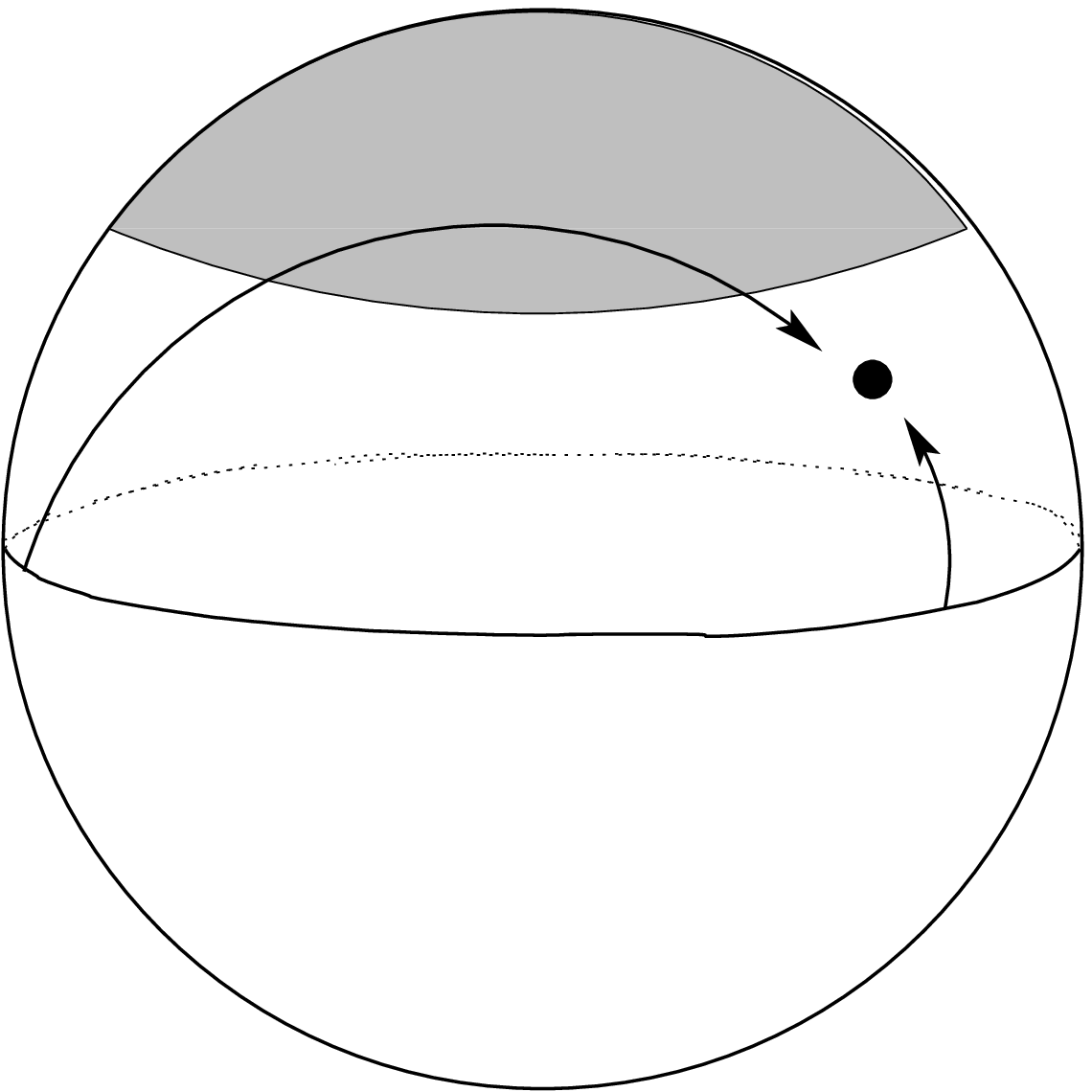,width=.23\textwidth}\hspace{1cm}
\epsfig{file=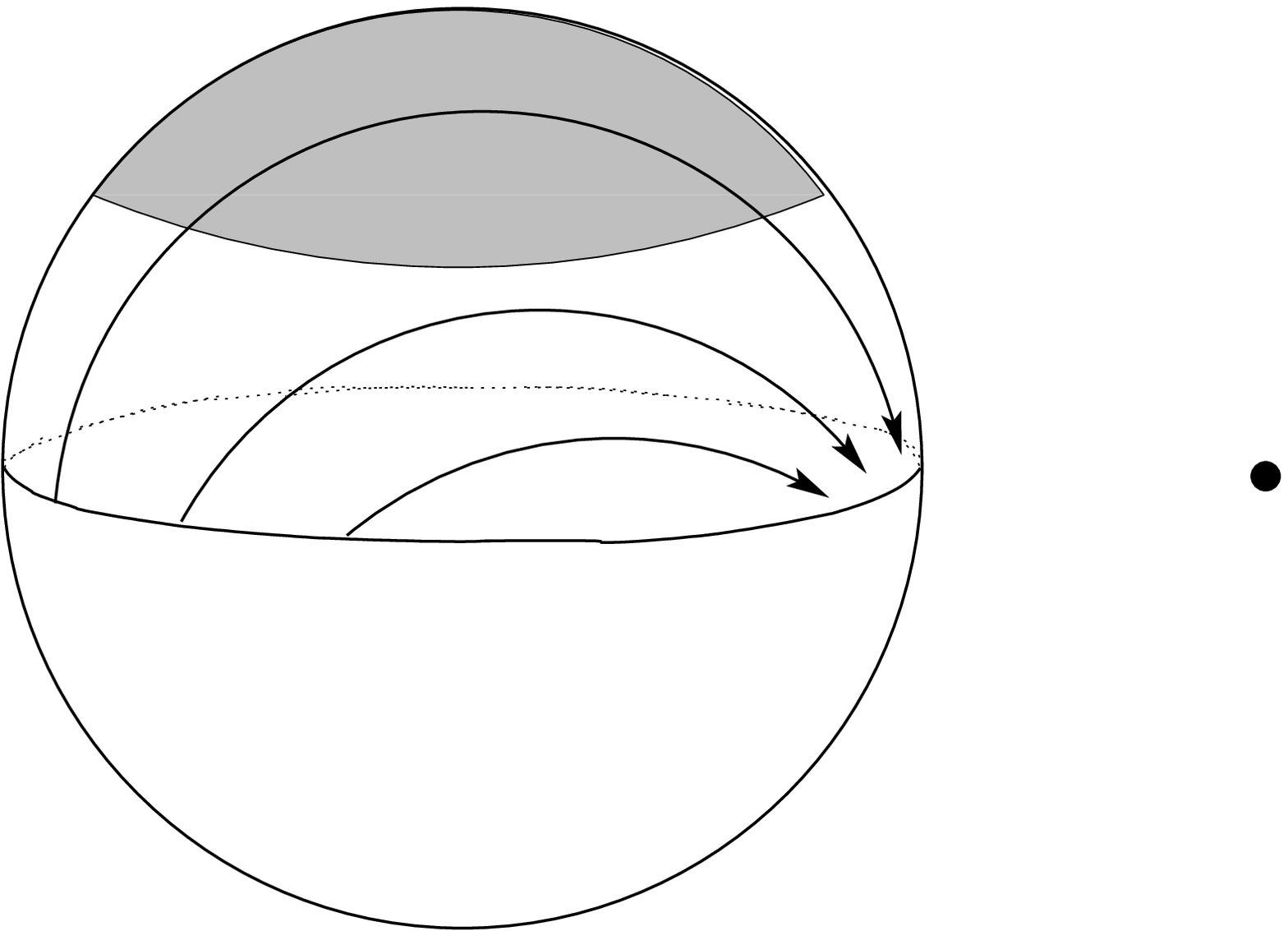,width=.33\textwidth}}
 \caption{\it Each cosmological solution is represented by a curve on the sphere.
In the left figure ``Rome", represented by the dot, is on the sphere and each curve is directed
from the equator towards ``Rome", which corresponds to a power-law solution to the equations of
motion. In the right figure ``Rome" is not on the sphere and each curve, again being directed
towards ``Rome", begins and ends on the equator. In this case ``Rome" is not a solution. The
accelerated expansion of the solution happens whenever the curve lies within the ``arctic circle".
This region is given by the shaded area.}
\end{figure}

In this work we will discuss systematically the accelerating phases of all 2-scalar cosmologies
with a single exponential potential by associating to each solution a trajectory on a two-sphere.
It turns out that all trajectories have the property that, when projected onto the equatorial
plane, they reduce to straight lines which are directed towards a point that we will call ``Rome".
Depending on the specific dilaton coupling of the potential, this point can be either on the sphere
or not. In the former case, it corresponds to a power-law solution for the scale factor, whereas in
the latter case, it is not a solution. We find that the accelerating phase of a solution is
represented by the part of the trajectory that lies within the ``arctic circle" on the sphere, see
figure 1. This enables us to calculate the expansion factors in a straightforward way for each of
the solutions.

In the second part of this work we show the higher-dimensional origin of a class of solutions.
Applying 3D group manifold reductions we consider the embedding in gauged supergravities with an
M-theory origin. The higher-dimensional origin of certain power-law
solutions is a (locally) Minkowskian space-time in $D \geq 6$ dimensions \cite{Gutperle:2002ai,Kruczenski:2002ap,Russo:2003ky}.

We extend our analysis to $n$-tuple exponential potentials with $n=2,3$ and $6$ that follow from
the group manifold reduction. In the case of double exponential potentials we compare our results
with those that follow from the flux compactification on a maximally symmetric space and S-branes.
In the triple exponential case our analysis suggests the existence of two exotic S($D-3$)-branes in
$D$ dimensions which have not been constructed before in the literature.

This paper is organised as follows. In sections \ref{Multi-Scalar}-\ref{Interpolating} we present,
under the assumptions stated, the most general $N$-scalar accelerating cosmology in 4 dimensions.
The accelerating phases of these cosmologies are discussed in section \ref{acceleration}. Their
equations of state and the one-scalar truncations are discussed in sections \ref{eos} and
\ref{1scalar}, respectively. In section \ref{M-theory} we discuss the higher-dimensional origin of
a class of the $D=4$ solutions. In section \ref{gmr} we explain the group manifold reduction and in
section \ref{truncation-single} we consider the special case of a single exponential potential. In
sections \ref{double-section}-\ref{triple} we consider the case of double and triple exponential
potentials and relate to other reductions and S-branes. The embedding into M-theory and gauged
maximal supergravities is discussed in section \ref{gsg}. Finally, in section 4 we comment on
inflation, the compactness of the group manifolds and exotic S-branes.

\section{ Accelerating Cosmologies in 4 dimensions}

\subsection{Multi-scalar-gravity with a Single Exponential Potential} \label{Multi-Scalar}

Our starting point is gravity coupled to $N$ scalars \cite{Copeland:1999cs} which we denote by
$(\varphi, \vec{\phi})$. We assume that the scalar potential consists of a single exponential term:
\begin{align}
 {\cal L}= \sqrt{-g} \,\Big[R-\tfrac{1}{2} (\partial\varphi)^2
-\tfrac{1}{2}(\partial\vec{\phi})^2-V(\varphi,\vec{\phi})\Big] \,,
 \qquad
 V(\varphi, \vec{\phi}) = \Lambda\, \exp(-\alpha \varphi - \vec{\beta} \cdot \vec{\phi})
 \,,
 \label{LagrangianN}
\end{align}
where we restrict\footnote{We make this choice in order to obtain quintessence and therefore
accelerating solutions.} to $\Lambda>0$. To characterise the potential we introduce the following
parameter:
\begin{align}
\Delta \equiv \alpha^2+|\vec{\beta}|^2-\frac{2\,(D-1)}{(D-2)}=\alpha^2+|\vec{\beta}|^2-3\quad
\text{for}\quad D=4.
\end{align}
This parameter, first introduced in \cite{Lu:1995cs}, is invariant under toroidal reductions.

The kinetic terms of the dilatons are invariant under $SO(N)$-rotations of $(\varphi, \vec{\phi})$.
However, in the scalar potential the coefficients $\alpha$ and $\vec{\beta}$ single out one
direction in $N$-dimensional space. Therefore the Lagrangian \eqref{LagrangianN} is only invariant
under $SO(N-1)$. The remaining generators of $SO(N)$ can be used to set $\vec{\beta} = 0$, in which
case only the scalar $\varphi$ appears in the scalar potential. Such a choice of basis leaves
$\Delta$ invariant.

Motivated by observational evidence, we choose a flat FLRW Ansatz. This basically means a spatially
flat metric that can only contain time-dependent functions. One can always perform a
reparametrisation of time to bring the metric to the following form:
\begin{align}
ds^2 &= -a(u)^{2\,\delta}\,du^2+a(u)^2\,dx_3^2 \label{FRW} \,,
\end{align}
for some $\delta$. In this paper we will choose $\delta$ as
follows\footnote{The non-cosmic time corresponds to the gauge in
which the lapse function $N\equiv \sqrt{- g_{tt}}$ is equal to the
square root of the determinant $\gamma$ of the spatial metric,
i.e.~$N=\sqrt{\gamma}$, whereas cosmic time corresponds to $N=1$.
We thank Marc Henneaux for a discussion on this point.}:
\begin{alignat}{6}
 & \text{Cosmic time:}& \quad \delta&=0 \,, \qquad u&=\tau \,, & \quad &\frac{da}{d\tau}&=\dot{a}\,, \label{cosmic}\\
 & \text{Non-cosmic time:} & \quad \delta&=3 \,, \qquad u&=t \,, & \quad &\frac{da}{dt}&=a'\,.\label{time}
\end{alignat}
As a part of the Ansatz, we also assume:
\begin{align}
\varphi = \varphi (u)\,, \qquad \vec{\phi} = \vec{\phi} (u) \,.
 \end{align}
For this Ansatz one can reduce the $N-1$ scalars $\vec{\phi}$ that do not appear in the potential to
one scalar by using their field equations as follows:
\begin{align}
\frac{d^2\,\vec{\phi}}{du^2} &= (\delta-3)\,\frac{d\,\log{a}}{du}\,\frac{d\,\vec{\phi}}{du} \qquad
\Rightarrow \qquad \frac{d\,\vec{\phi}}{du} = \vec{c}\,a^{\delta-3}\,,
\end{align}
where $\vec{c}$ is some constant vector. The only influence of the $N-1$ scalars comes from their
total kinetic term:
\begin{align}
 \Big| \frac{d\,\vec{\phi}}{du} \Big|^2 &= |\vec{c}|^2\,a^{2\,\delta-6}\,.
\end{align}
Therefore, from the metric point of view, there is no difference between $N=2$ and $N>2$ scalars
(under the restriction of a single exponential potential). The truncation of the system
\eqref{LagrangianN} to one scalar corresponds to setting $\vec{c} = 0 $.\\

To summarise, we will be using the following Lagrangian:
\begin{align}
 {\cal L}= \sqrt{-g} \,\Big[R-\tfrac{1}{2}(\partial\varphi)^2
-\tfrac{1}{2}(\partial\phi)^2-V(\varphi)\Big] \,,
 \qquad
 V(\varphi) = \Lambda\, \exp(-\alpha \varphi ) \,,
 \label{Lagrangian2}
\end{align}
with $\Lambda>0$ and we choose the convention $\alpha\geq0$. From now on
we will use $\Delta = \alpha^2 -3$ instead of $\alpha$.

In the next two subsections we will first discuss the critical points corresponding to the system
(\ref{Lagrangian2}) and then the solutions that interpolate between these critical points. We will
use cosmic time \eqref{cosmic} when discussing the critical points in section \ref{Critical} and
non-cosmic time \eqref{time} when dealing with the interpolating solutions in section
\ref{Interpolating}.

\subsection{Critical Points} \label{Critical}

It is convenient to choose a basis for the fields, such that they parametrise a 2-sphere. In this
basis we will find that all constant configurations (critical points) correspond to power-law
solutions for the scale factor $a(\tau)\sim \tau^p$ for some $p$. By studying the stability of
these critical points \cite{Halliwell:1987ja, Copeland:1998et} one can deduce that there exist
interpolating solutions which tend to these points in the far past or the distant future. We will
actually be able to draw these interpolating solutions without having to do any stability analysis. \\

We begin by choosing the flat FLRW Ansatz \eqref{FRW} in cosmic time:
\begin{align}
ds^2&=-d\tau^2+a(\tau)^2\,(dx^2+dy^2+dz^2)\,.
\end{align}
The Einstein equations for the system \eqref{Lagrangian2} with this Ansatz become:
\begin{align}
H^2&=\tfrac{1}{12}\,(\dot\varphi^2+\dot\phi^2)+\tfrac{1}{6}\,{V}\,,
\label{Friedmann}\\
\dot{H} &=-\tfrac{1}{4}\,(\dot{\varphi}^2+\dot{\phi}^2)\,, \label{Acceleration}
\end{align}
where $H \equiv \dot{a}/a$ is the Hubble parameter and the dot denotes differentiation
w.r.t.~$\tau$. Equations \eqref{Friedmann} and \eqref{Acceleration} are usually referred to as the
Friedmann equation and the acceleration equation, respectively. The scalar equations are:
\begin{align}
\ddot\varphi &=-3\,H\,\dot\varphi+\sqrt{\Delta+3}\,\,V\,,\qquad\ddot\phi=-3\,H\,\dot\phi\,.
\label{scalars}
\end{align}
We define the following three variables:
\begin{align}
x&=\frac{\dot\varphi}{\sqrt{12}\,H}\,,\qquad y=\frac{\dot\phi}{\sqrt{12}\,H}\,,\qquad
z=\frac{\sqrt{V}}{\sqrt{6}\,H}\,.
\end{align}
In these variables the Friedmann equation \eqref{Friedmann} becomes the defining equation of a
2-sphere \cite{Copeland:1998et,Guo:2003rs}:
\begin{align}
x^2+y^2+z^2&=1\,. \label{sphere}
\end{align}
This means that we can think of solutions as points or
trajectories on a globe. It turns out that cosmological solutions
are either eternally expanding (i.e.~$H>0$) or eternally
contracting ($H<0$), but cannot have an expanding phase and then a
contracting phase (or vice-versa). Since we are only interested in
expanding universes, we will only be concerned with the upper
hemisphere (i.e.~$z>0$). In terms of $x$ and $y$ the scalar
equations become:
\begin{align}
\frac{\dot{x}}{H}&=-3\,z^2\,(x-\sqrt{1+\Delta/3})\,,\label{x}
\\ \frac{\dot{y}}{H}&=-3\,z^2\,y\,. \label{y}
\end{align}
We can rewrite the acceleration equation \eqref{Acceleration} as
follows:
\begin{align}
\frac{\dot{H}}{H^2}&=-3\,(x^2+y^2)\,. \label{Acceleration'}
\end{align}
If we now solve for the critical points ($\dot{x}=0, \dot{y}=0$), we can then integrate
\eqref{Acceleration'} twice and obtain the following power-law solutions for $a(\tau)$
\cite{Copeland:1999cs}:
\begin{align}
a(\tau)\sim \tau^p\,,\quad \text{where} \quad p=\frac{1}{3\,(x_c^2+y_c^2)}\,,
\end{align}
and the following solutions for the scalars:
\begin{align}
\varphi&=\sqrt{12}\,p\,x_c\,\log(\tau)+\text{constant}\, .
\end{align}
We thus find the following critical points:
\begin{itemize}
\item \underline{The equator}:
\begin{align}
z=0, \quad x^2+y^2=1.
\end{align}
Every point on the equator of the sphere is a critical point with power-law behaviour $a\sim
\tau^{1/3}$. \item \underline{``Rome"}:
\begin{align}
x=\sqrt{1+\Delta/3}\,, \quad y=0\,, \quad z=\sqrt{-\Delta/3}\,.
\end{align}
This critical point yields a power-law behaviour of the form (we ignore here irrelevant constants
that rescale time)
\begin{equation}
a\sim\tau^{1/(\Delta+3)}\ \ {\rm for}\ \ -3 < \Delta < 0\, ,\hskip 1truecm
a\sim e^\tau\ \ {\rm for}\ \ \Delta = -3.
\end{equation}
Note that the greater $\Delta$ is, the further ``Rome" gets pushed
towards the equator, and for $\Delta =0$ it is on the equator.
\end{itemize}
Although the equatorial points (a.k.a.~kinetic-dominated solutions) do solve
\eqref{sphere}-\eqref{Acceleration'} as critical points, they are not proper solutions of
\eqref{Friedmann}-\eqref{scalars} in terms of the fundamental fields, since $z=0$ would imply that
$V=0$, which is impossible for $\Lambda\neq0$ unless $\varphi$ is infinite at all times. However,
these points will be interesting to us, as they will give information about the asymptotics of the
interpolating solutions.

In contrast to the equator, the ``Rome" critical point is a
physically acceptable solution of the system, provided it is well
defined on the globe (i.e.~$\Delta<0$). In the case where
$\Delta=-3$ it becomes De Sitter (i.e.~$a\sim e^\tau$), as one
would expect, since $V=\Lambda$.

Besides these critical points there are other solutions, which are
not points but rather trajectories. In fact, we can already
determine their shapes. Dividing \eqref{x} and \eqref{y} we obtain
the following:
\begin{align}
\frac{dy}{dx}&=\frac{y}{x-\sqrt{1+\Delta/3}}\,.\label{diffline}
\end{align}
Integrating this we get the following relation between $x$ and $y$:
\begin{align}
y&=C\,(x-\sqrt{1+\Delta/3})\,, \label{line}
\end{align}
where $C$ is an arbitrary constant\footnote{Since $C$ is finite
one might think that this excludes the line defined by
$x=\sqrt{1+\Delta/3}$. However, that line can be obtained by
taking the inverse of \eqref{diffline} and solving for $x$ as a
function of $y$.}. This relation tells us that if we project the
upper hemisphere onto the equatorial plane, in other words, if we
view the sphere from above, any solution to the equations of
motion must trace out a straight line that lies within the circle
defined by $x^2+y^2=1$ and has a $y$-intercept at
($x=\sqrt{1+\Delta/3}, y=0$). From now on, we will refer to that
point as ``Rome"\footnote{Note that we have extended our
definition of ``Rome": only if ``Rome" is on the globe ($\Delta
<0$) is it equal to the critical point discussed before.}. Notice
that all lines intersect at ``Rome" independently of whether it is
on the globe ($\Delta<0$), right on the equator ($\Delta=0$) or off
the globe ($\Delta>0$). These lines can only have critical points
as end-points. So each line is a solution, which interpolates
between two power-law solutions. In a similar, yet physically
inequivalent context, such a line was found in
\cite{Coley:1997nk}.

\begin{figure}[h]
\centerline{\epsfig{file=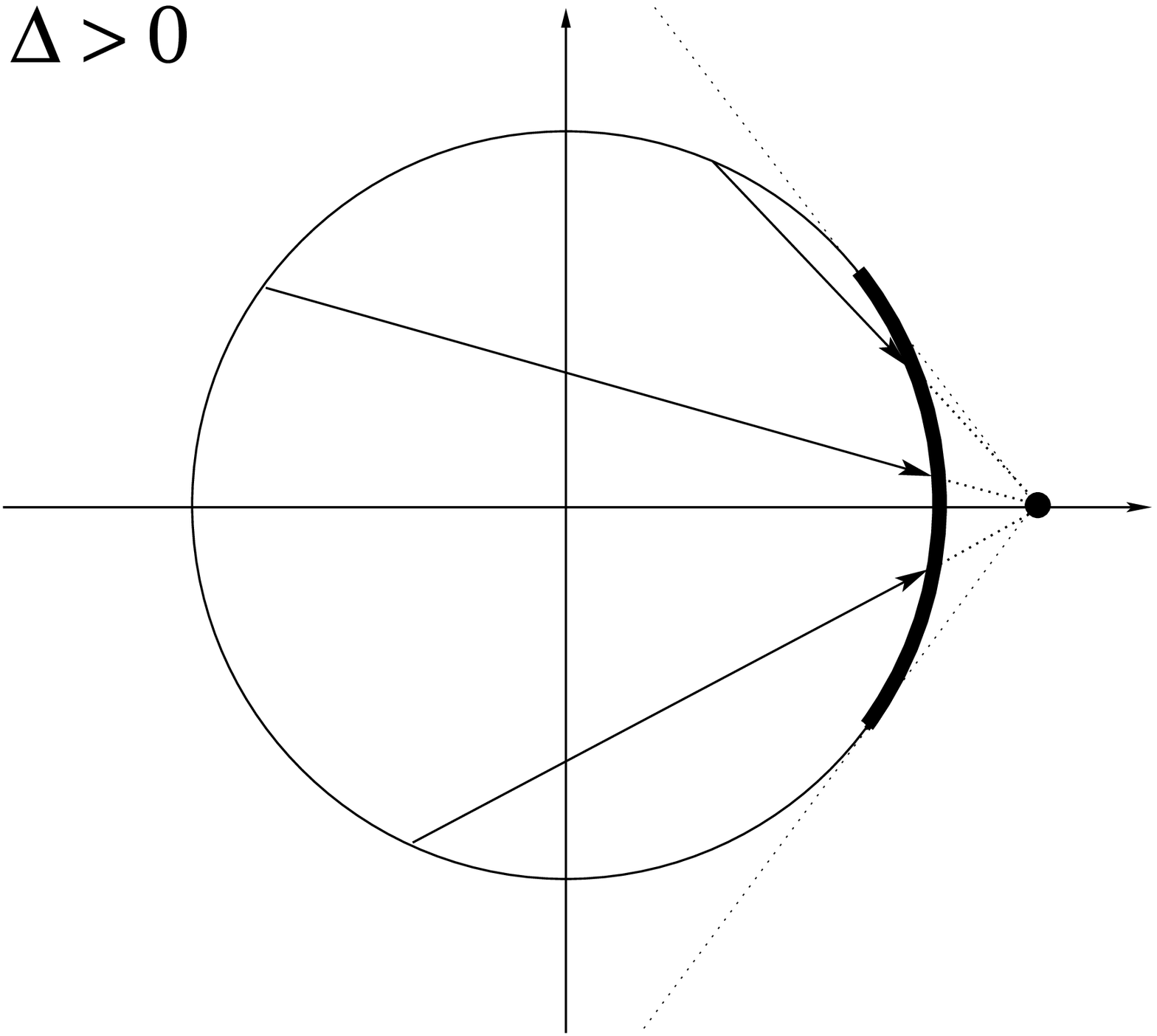,width=.33\textwidth}
 \epsfig{file=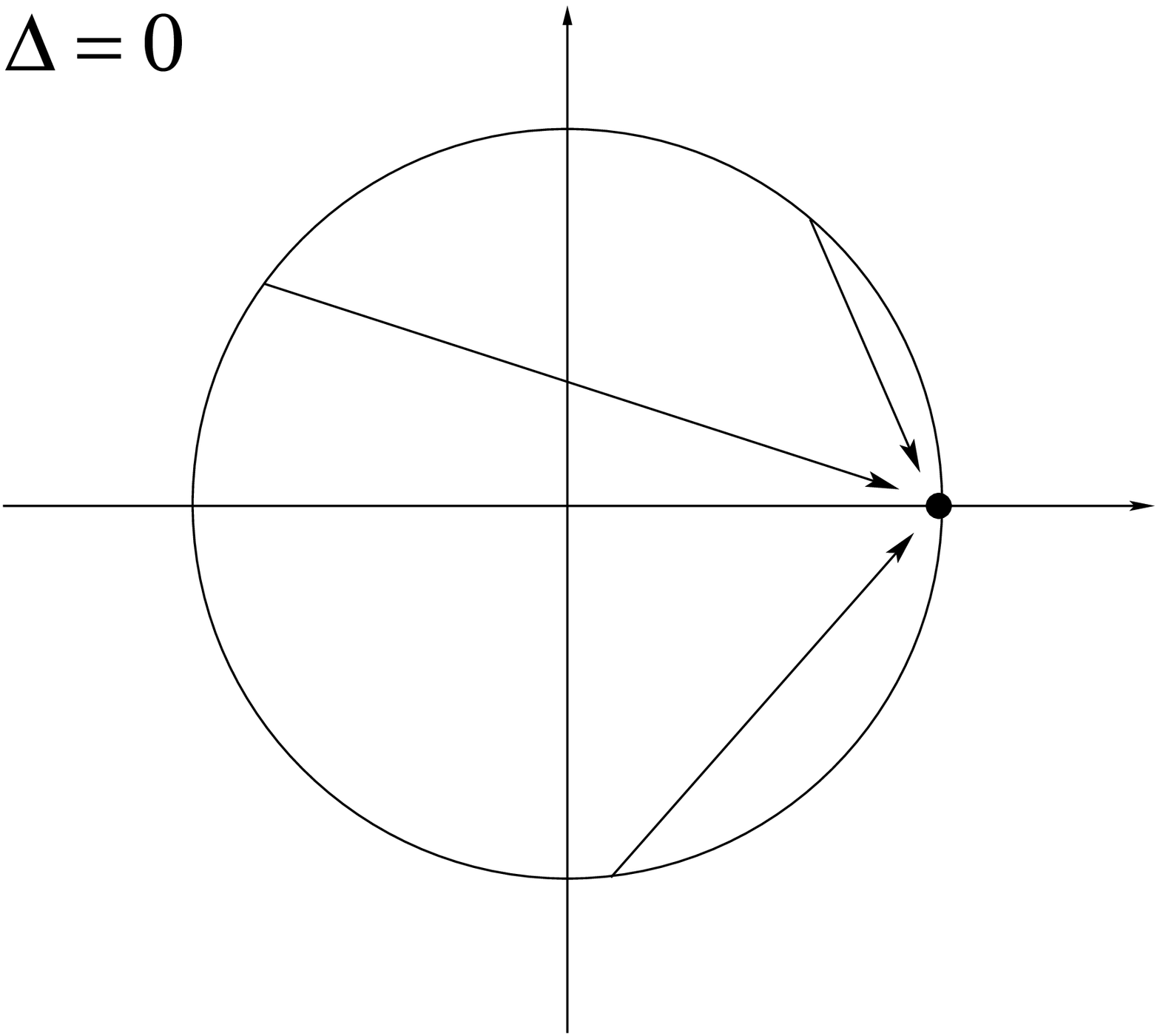,width=.33\textwidth} \epsfig{file=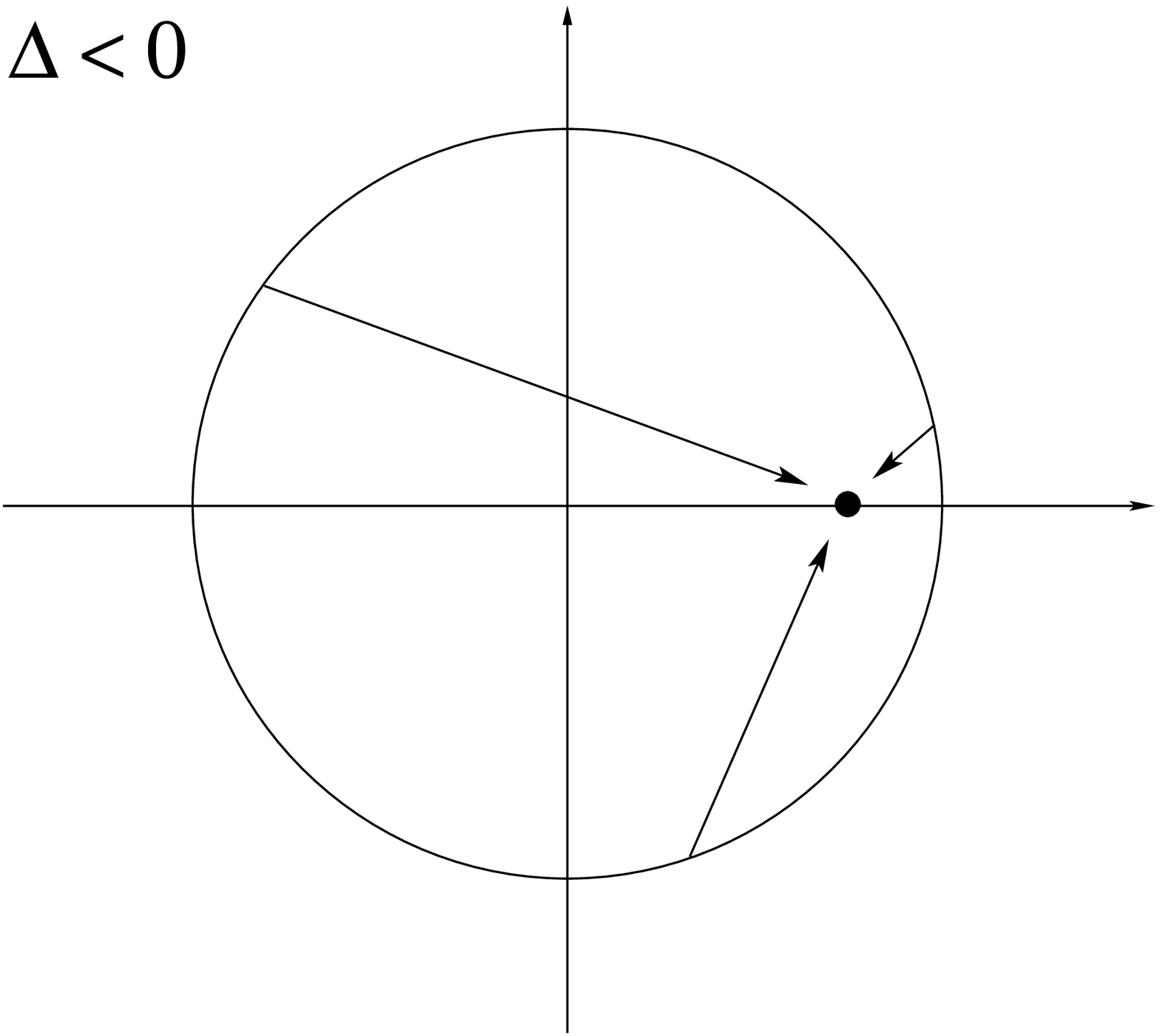,width=.33\textwidth}}
 \caption{\it The solutions represented as straight lines in the $(x,y)$-plane for $\Delta>0$ where
 ``Rome" is not on the sphere, $\Delta=0$ where ``Rome" is on the equator and $-3 \le \Delta<0$ where ``Rome" is on the
 sphere. The thick arc in the left figure represents the attracting portion
of the equator \cite{Billyard:1999dg}.} \label{interpfig}
\end{figure}

Now that we know the shapes of the trajectories, let us figure out their time-orientations. By
looking at \eqref{x} we realise that the time derivative of $x$ is positive when
$x<\sqrt{1+\Delta/3}$ and negative when $x>\sqrt{1+\Delta/3}$. This tells us that {\it all roads
lead to Rome}. Figure \ref{interpfig} illustrates this for the cases where ``Rome" is off the
globe, right on the equator or on the globe.

One can also determine the orientations of the trajectories by analysing the stability of the
critical points. One will find that whenever ``Rome" is on the globe (i.e.~$\Delta=0$ and
$\Delta<0$), it is stable (i.e.~an attractor), and the points on the equator are all unstable
(i.e.~repellers), except for ``Rome" when $\Delta=0$. In the case where ``Rome" is off the globe
(i.e.~$\Delta>0$), the equator splits up into a repelling and an attracting region. The attracting
region turns out to be the portion of the equator that can ``see'' ``Rome". In other words, any
point on the equator that can be joined to ``Rome" by a straight line such that the line does not
intersect the equator again before reaching ``Rome" is attracting. To summarise, for $\Delta>0$,
all points on the equator with $x>\sqrt{3/(\Delta+3)}$ are attracting, and the rest are repelling.
In the first illustration of figure \ref{interpfig}, the attracting portion of the equator is
depicted by the thick arc.

\subsection{Interpolating Solutions} \label{Interpolating}

To solve the equations of motion, it is convenient to use the FLRW Ansatz \eqref{FRW} in non-cosmic
time:
\begin{align}
ds^2 &= -a(t)^6\,dt^2+a(t)^2\,dx_3^2\,.
\end{align}
Substituting this Ansatz in the Einstein equations yields
\begin{align}
F^2 &= \tfrac{1}{3}\,F'+\tfrac{1}{12}\,(\phi'^2+\varphi'^2)\,,
\label{E1}\\
F' &= \tfrac{1}{2}\,V\,a^6\,, \label{E2}
\end{align}
where $F=a'/a$ is a Hubble parameter-like function, and the prime denotes differentiation w.r.t.~$t$. The equations for the scalars are:
\begin{align}
\phi'' &= 0\,,\qquad \varphi'' = \sqrt{\Delta+3}\,V\,a^6 \label{varphi}\,.
\end{align}
Combining \eqref{varphi} and \eqref{E2} gives the following solutions for the scalars:
\begin{align}
\varphi&=2\,\sqrt{\Delta+3}\,\log(a)+a_1\,t+b_1\,,\qquad \phi=a_2\,t+b_2\,. \label{solscalars}
\end{align}
By substituting this into equation \eqref{line} we can deduce that the slope of the line is given
by $C=a_2/a_1$. Substituting the scalars into \eqref{E1} and \eqref{E2} we are now left with the
following two equations:
\begin{align}
F' &= -\Delta\,F^2-\sqrt{\Delta+3}\,a_1\,F-\tfrac{1}{4}\,(a_1^2+a_2^2)\\
&= \tfrac{1}{2}\,\Lambda\,e^{-\sqrt{\Delta+3}\,(b_1+a_1\,t)}\,a^{-2\,\Delta}\,. \label{F}
\end{align}
Keeping in mind that $F'$ must be positive due to \eqref{F} we can now solve for $F$ in the three
different cases where $\Delta$ is positive, zero and negative. We can then easily find $a(t)$. We
will choose $b_1$ (the constant part of $\varphi$) such that all solutions for $a(t)$ have a
proportionality constant of $1$, which does not affect the cosmological properties of the
solutions. The integration constants appearing in the solutions are defined as follows:
\begin{align}
 c_1 = \frac{-\sqrt{\Delta+3}\,a_1}{2\,\Delta}\,, \quad c_2 =
\frac{\sqrt{3\,a_1^2-\Delta\,a_2^2}}{2}\,, \quad d_1 =-\frac{a_1^2+a_2^2}{4\,\sqrt{3}\,a_1}\,,
\quad d_2 = -\sqrt{3}\,a_1\,.
\end{align}
Below we present the solutions \cite{Gavrilov:1994sv,Aguirregabiria:2000hx} and their late- and
early-time asymptotic behaviours (we give the latter without any irrelevant constants that rescale
time):
\begin{enumerate}
\item \underline{$\Delta>0$}:
\begin{align}
a(t)&= e^{c_1\,t}\,\cosh{(c_2\,t)}^{1/\Delta}\,, \qquad \text{for} \quad -\infty<t<+\infty\,,
\end{align}
The positivity of $F'$ requires $a_1$ to be negative, and it also imposes the following constraint:
\begin{align}
\Big(\frac{a_2}{a_1}\Big)^2<\frac{3}{\Delta}\,. \label{constraint}
\end{align}
This solution corresponds to a generic line on the first illustration in figure \ref{interpfig}. It
starts on the equator somewhere to the left of $x=\sqrt{3/(\Delta+3)}$, then moves in the direction
of ``Rome", but ends on the equator on the right-hand side. Note that the constraint
\eqref{constraint} is simply the requirement that the slope of the line is bounded from above and
from below such that the line actually intersects the sphere. We can confirm this asymptotic
behaviour of the solution by converting to cosmic time \eqref{cosmic} for $t\rightarrow -\infty$
and $t\rightarrow +\infty$ with the relation $a(t)^3\,dt=d\tau$:
\begin{alignat}{3}\nonumber
t&\rightarrow -\infty\,,& \qquad \tau &\rightarrow 0\,,& \qquad
&a\rightarrow e^{t}\sim\tau^{1/3}\,,\\
t&\rightarrow +\infty\,,& \qquad \tau &\rightarrow +\infty\,,& \qquad &a\rightarrow
e^{t}\sim\tau^{1/3}\,.
\end{alignat}
\item \underline{$\Delta=0$}:
\begin{align}
a(t)&= e^{d_1\,t}\,\exp{(e^{d_2\,t})}\,, \qquad \text{for} \quad -\infty<t<+\infty\,.
\end{align}
The positivity of $F'$ requires $a_1$ to be negative. This
corresponds to a line on the second illustration in figure
\ref{interpfig}. It starts on the equator and reaches
``Rome"\footnote{In this case, ``Rome" is again attracting,
however to see that, one must perform the stability analysis by
going to second order perturbation. The first order vanishes,
which means that the interpolating trajectory approaches ``Rome"
more slowly than in the cases where $\Delta<0$. }, which is also
on the equator. Its asymptotic behaviour goes as follows:
\begin{alignat}{3}\nonumber
t&\rightarrow -\infty\,,& \qquad \tau &\rightarrow 0\,,& \qquad
&a\rightarrow e^{t}\sim\tau^{1/3}\,,\\
t&\rightarrow +\infty\,,& \qquad \tau &\rightarrow +\infty\,,& \qquad &a\rightarrow
e^{e^{t}}\sim\tau^{1/3}\,.
\end{alignat}
To find the late-time behaviour of $a$ in cosmic time one must
realise the following two facts: First, $a(t)\sim\exp(e^t)$ for
$t\rightarrow\infty$. Second, in this limit, $a'\sim a$ and
therefore $a$ behaves like a normal exponential.
\item
\underline{$-3\le\Delta<0$}:
\begin{align}
a(t)&= e^{c_1\,t}\,\sinh{(-c_2\,t)}^{1/\Delta}\,,\qquad \text{for}\quad -\infty<t<0\,.\label{sinh}
\end{align}
This solution corresponds to any line on the third illustration in
figure \ref{interpfig}. It starts at any point on the equator and
ends at ``Rome". This is reflected in the asymptotics as follows:
\begin{alignat}{4}\nonumber
t&\rightarrow -\infty\,,& \qquad \tau &\rightarrow 0\,,& \qquad
&a\rightarrow e^{t}\sim\tau^{1/3}\,,\\
t&\rightarrow 0\,,& \qquad \tau &\rightarrow +\infty\,,& \qquad
&a\rightarrow(-t)^{1/\Delta}\sim\tau^{1/(\Delta+3)}\,\quad &&\text{for}\quad\Delta>-3 \,, \\
&&&&&\hskip 2.25truecm \sim e^\tau\,&&\text{for}\quad\Delta=-3\, . \notag
\end{alignat}

There is one more solution for $-3\le\Delta<0$. If we set $a_1=a_2=0$ we find:
\begin{align}
a(t)&= (-t)^{1/\Delta}\,\qquad \text{for}\quad-\infty<t<0\,.\label{Romenc}
\end{align}
This solution corresponds to the ``Rome" solution itself. For $-3<\Delta<0$ the conversion to
cosmic time is the following:
\begin{align}
a\sim \tau^{1/(\Delta+3)}\,.
\end{align}
Notice, however, that in the case where $\Delta=-3$, the ``Rome" solution \eqref{Romenc} and
therefore the late-time asymptotics of \eqref{sinh} have a different conversion to cosmic time,
namely:
\begin{align}
a\sim(-t)^{1/\Delta}\sim e^\tau\,,
\end{align}
which we recognize as the De Sitter solution, in agreement with
the fact that we have $V=\Lambda$.
\end{enumerate}

The interpolating solutions above are given in non-cosmic time, which as mentioned is related to
cosmic time by
\begin{align}
d\tau=a(t)^3\,dt\,.
\end{align}
Integrating this equation yields hypergeometric functions for a generic interpolating solution,
which we can not invert to get the scale factor as a function of cosmic time. However, it is
possible to get interpolating solutions in cosmic time for negative $\Delta$ when the following
constraint on the constants holds:
\begin{align}
\Big(\frac{a_2}{a_1}\Big)^2= 12\,\frac{\Delta+\tfrac{9}{4}}{(2\Delta+3)^2}\,,
\end{align}
which can only be fulfilled for $-9/4\leq \Delta<0$. The relation between the two time coordinates
is
\begin{align}
\tau=\frac{2^{-3/\Delta}}{2 c_2}\,\frac{\Delta}{3+\Delta}\,(e^{2 c_2 t}-1)^{(3+\Delta)/\Delta}\,,
\end{align}
and the scale factor in cosmic time becomes
\begin{align}
a(\tau)=\Big(k_1\,\tau^{3/(3+\Delta)}+k_2\,\tau\Big)^{1/3}\,,
\end{align}
where $k_1=(2/c_1)^{3/\Delta}$ and $k_2=k_1 \,c_1\, (2\Delta+3)/(18+6\Delta)$. From this solution,
the asymptotic power-law behaviours are easily seen. The special one-scalar case, corresponding to
$\Delta=-9/4$, was found in \cite{Burd:1988ss}.

\subsection{Acceleration} \label{acceleration}

\begin{figure}[h]
\centerline{\epsfig{file=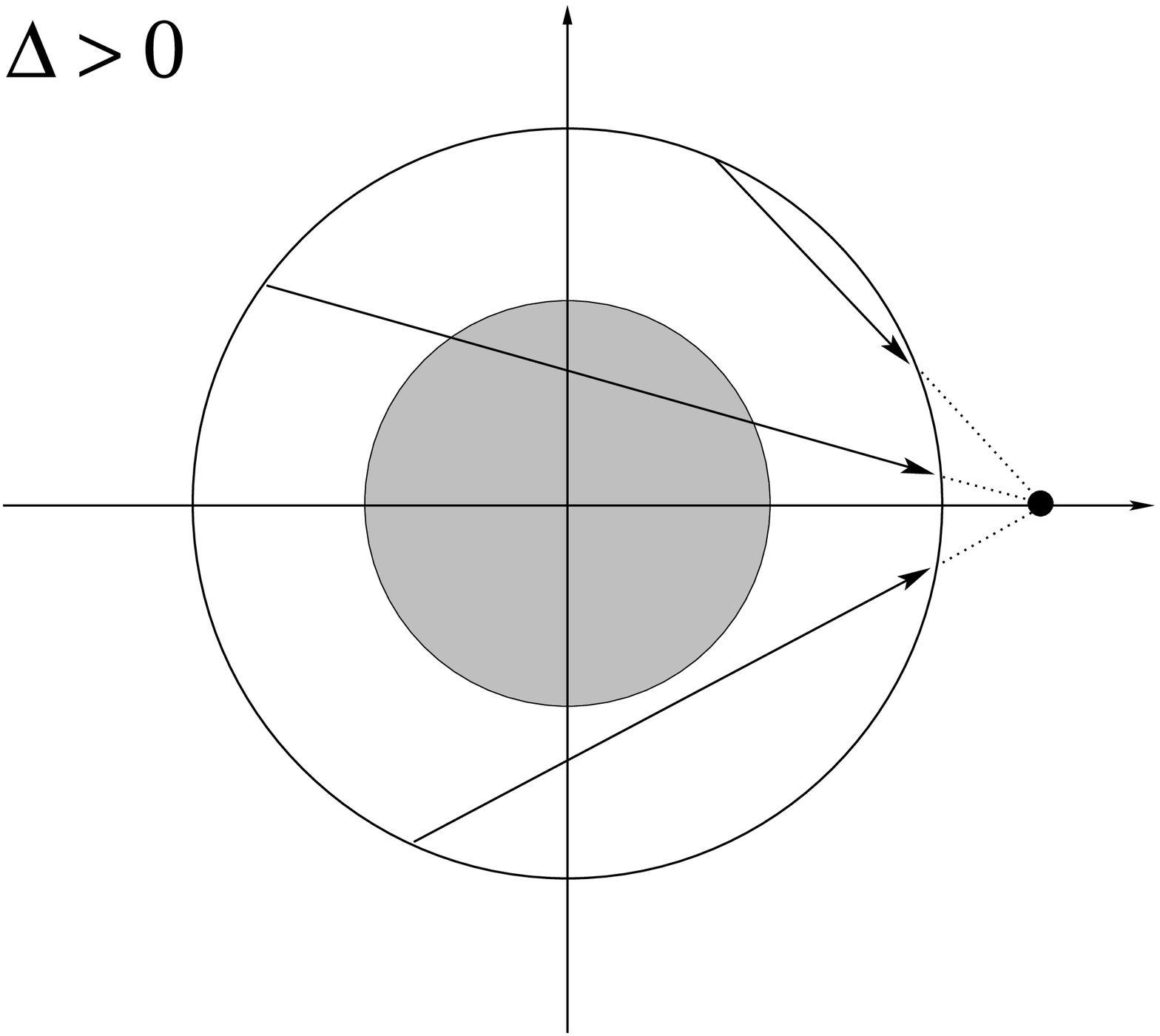,width=.33\textwidth}
 \epsfig{file=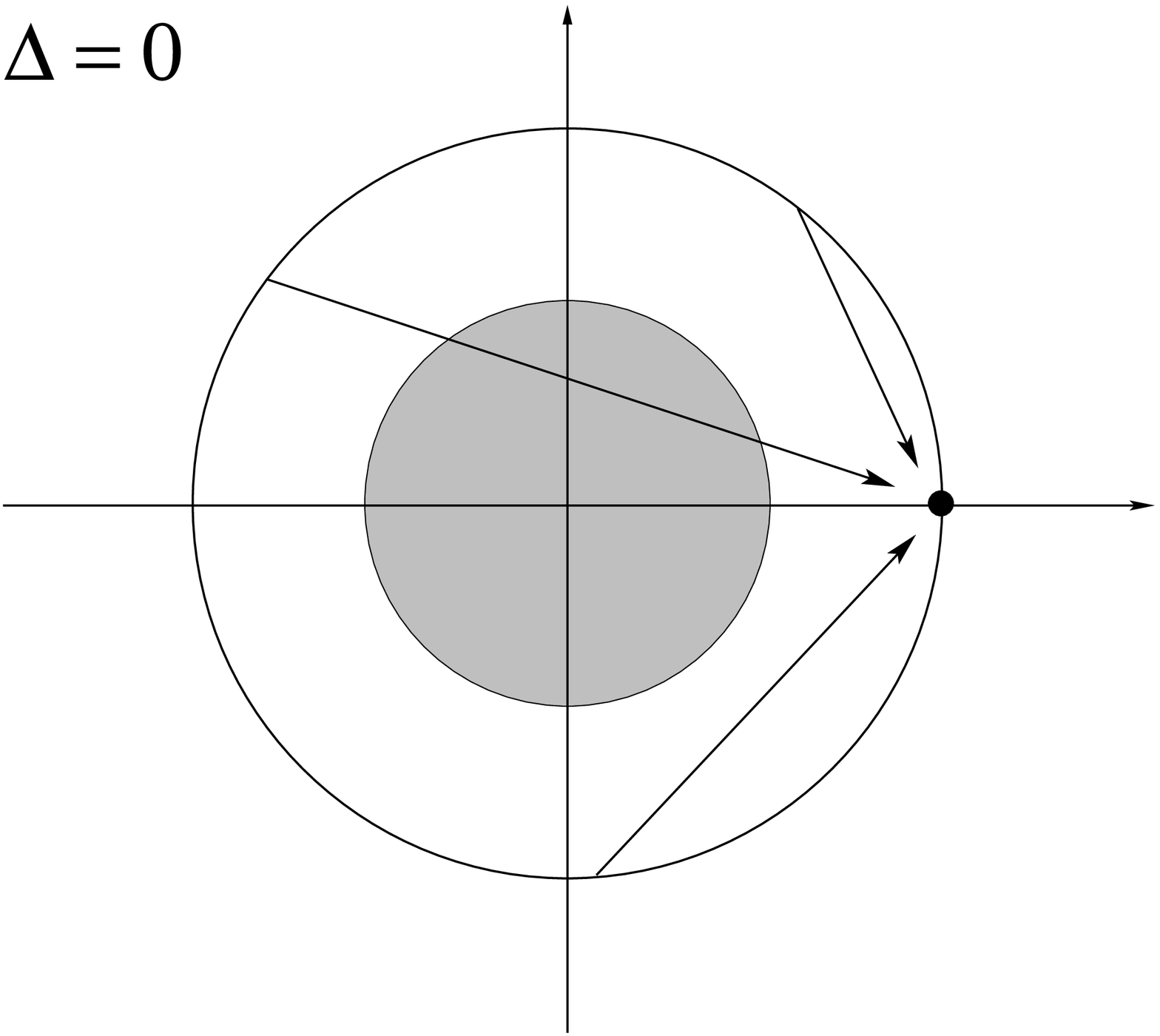,width=.33\textwidth} \epsfig{file=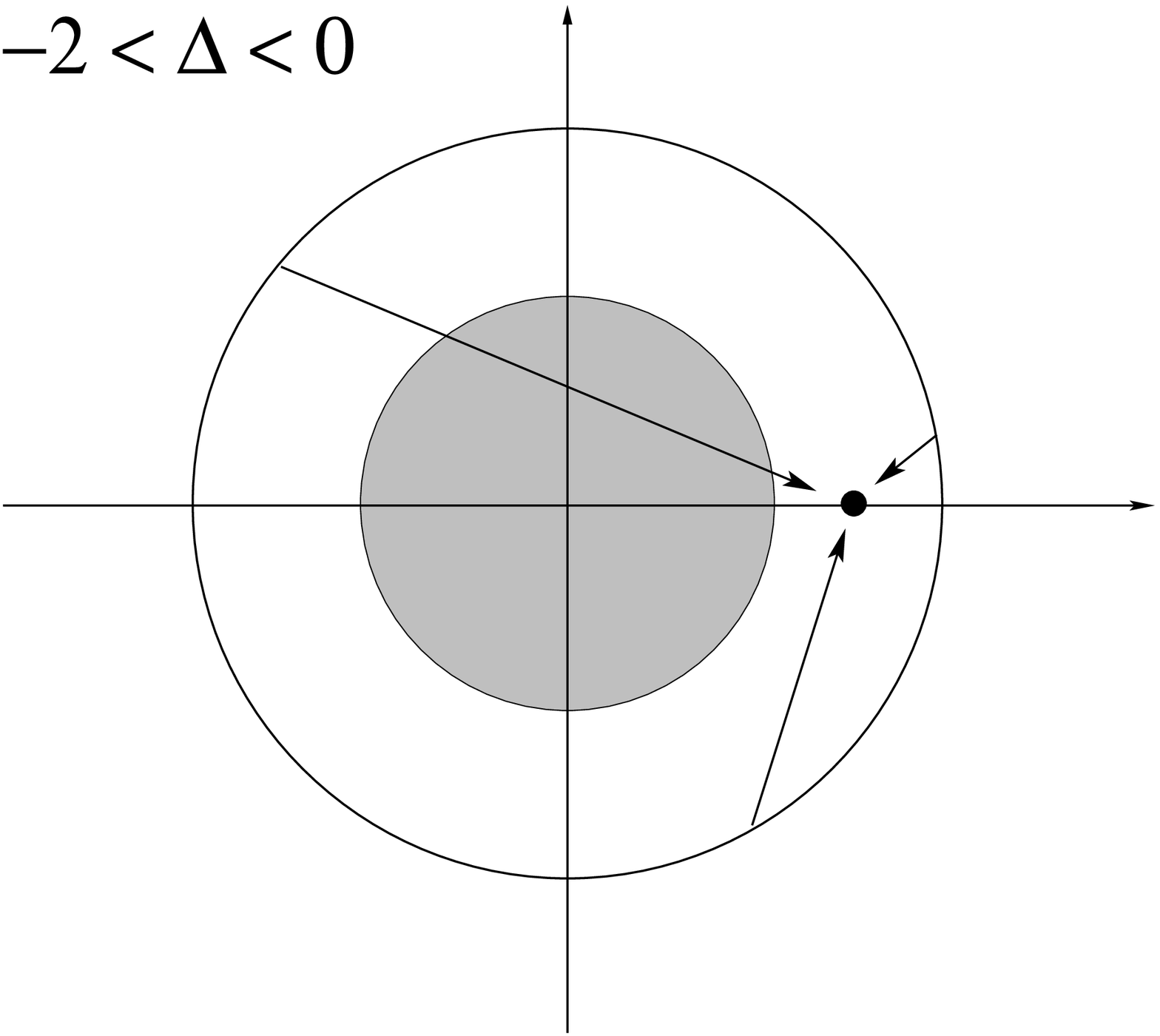,width=.33\textwidth}}
 \centerline{\epsfig{file=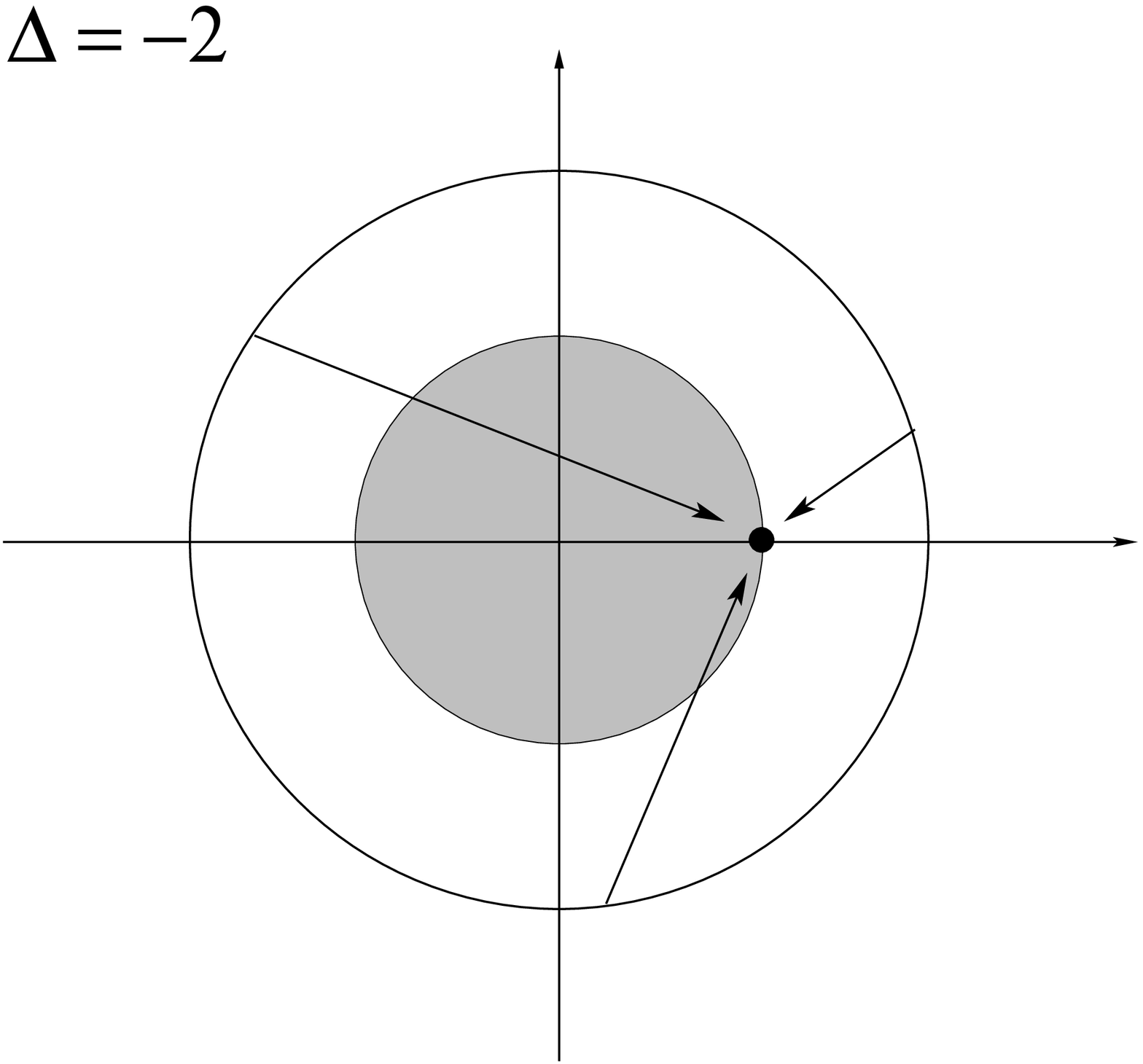,width=.33\textwidth}
 \epsfig{file=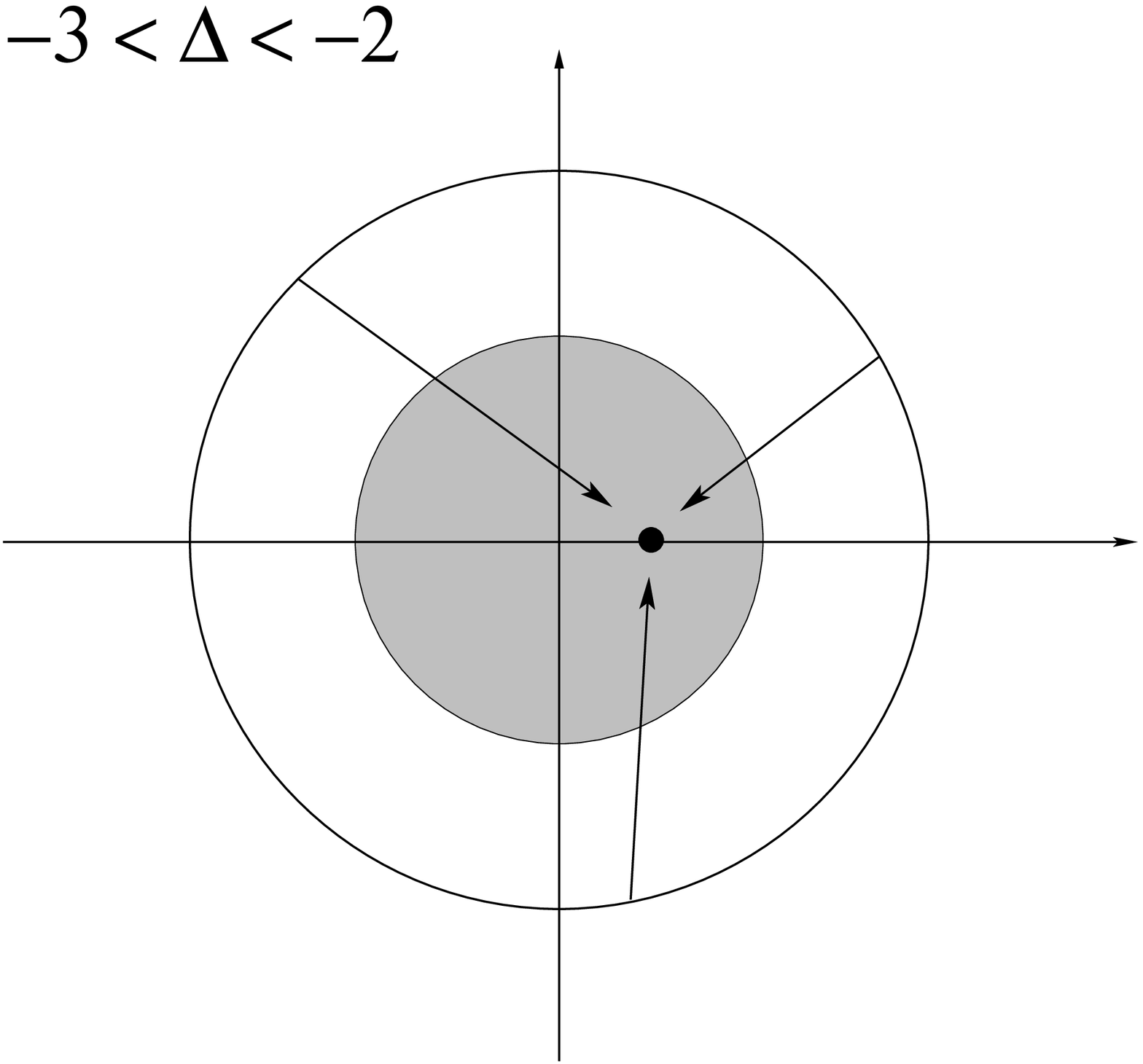,width=.33\textwidth} \epsfig{file=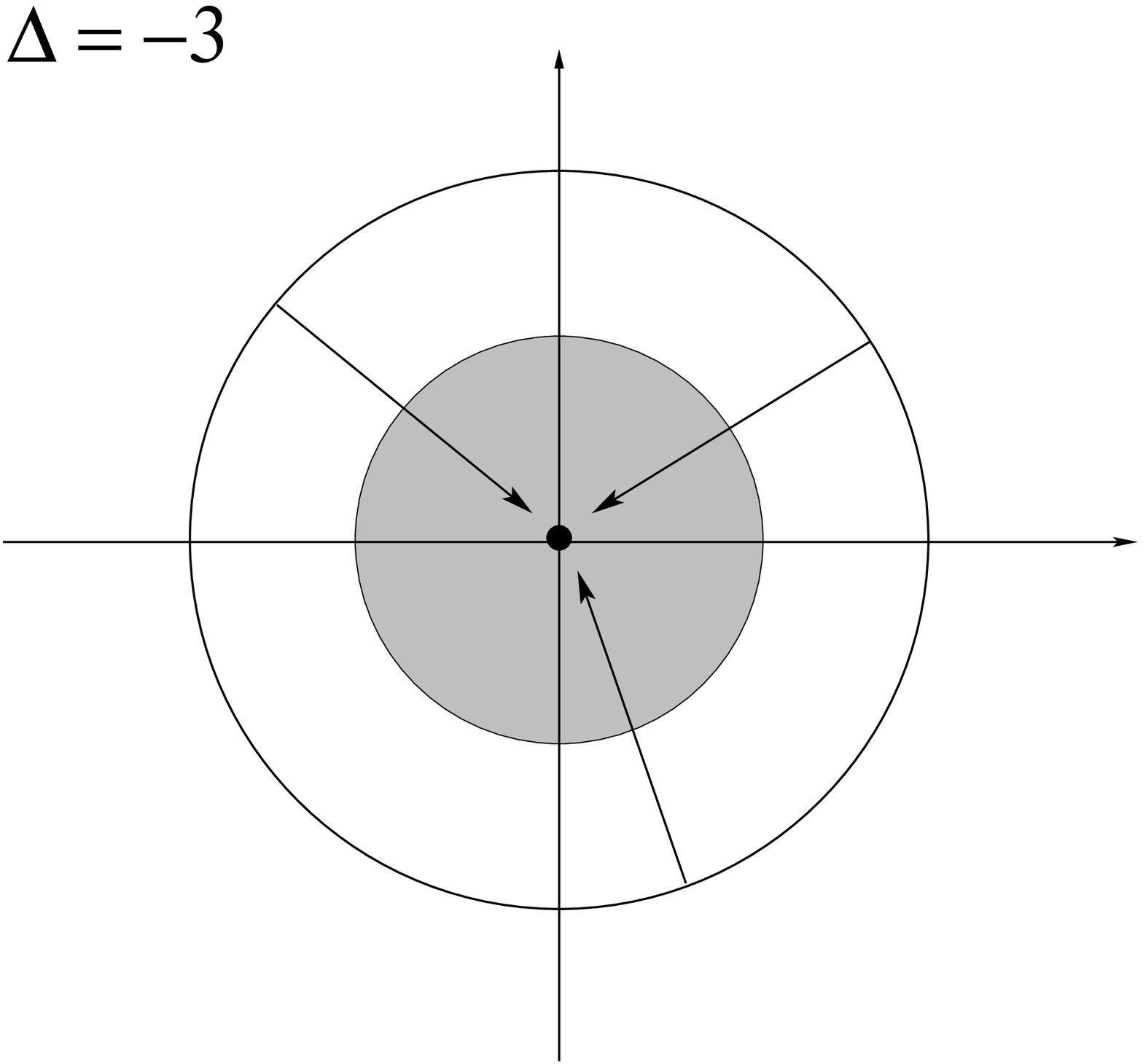,width=.33\textwidth}}
 \caption{\it The solutions represented as straight lines in the $(x,y)$-plane for $\Delta>0$ where
 ``Rome" is not on the sphere, $\Delta=0$ where ``Rome" is on the equator and $-3\le\Delta<0$ where ``Rome" is on the sphere.
 The inner circle corresponds to the ``arctic circle", and solutions are accelerating when they enter the shaded
 area. The lower part of the figure corresponds to the cases where
 ``Rome" is lying on the ``arctic circle", $\Delta=-2$, inside the ``arctic circle", $
-3<\Delta<-2$ and on the North Pole,
 $\Delta=-3$.}\label{lines}
\end{figure}

In this section we will investigate under which conditions ``Rome" and the interpolating solutions
represent an accelerating universe, i.e.~under which conditions we find quintessence. This can be
given a nice pictorial understanding in terms of the 2-sphere. We will show that quintessence
appears when the trajectory enters the region bounded by an ``arctic circle". This is summarised in
figure \ref{lines}.

An accelerating universe is defined by $\ddot{a}/a>0$. The existence of the ``arctic circle" in
connection to acceleration can now easily be determined. Assuming an expanding universe and using
\begin{align}
\frac{\ddot{a}}{a}=\dot{H}+H^2\,,
\end{align}
as well as \eqref{Acceleration'}, we see that the condition for acceleration is equivalent
to\footnote{A similar inequality was given in \cite{Coley:1997nk} for the one-scalar case, and in
terms of the scalars and the potential in \cite{Aguirregabiria:2000hx} for the multi-scalar case.}
\begin{align}\label{arctic}
z^2>\frac{2}{3}\,,\quad{\rm i.e.}\quad x^2+y^2<\frac{1}{3}\,,
\end{align}
which exactly yields an ``arctic circle" as the boundary of the region of acceleration. The straight
line representing the exact solution is parametrised by the constants $a_1$ and $a_2$ as found in
the previous section. From \eqref{arctic} and \eqref{line} it then easily follows that the
condition for acceleration leads to the following restriction for the slope of the line:
\begin{align}
\Big(\frac{a_2}{a_1}\Big)^2(2+\Delta) <1\,.\label{acca}
\end{align}
This condition is always fulfilled when $\Delta \leq -2$ and otherwise there is an interval of
values for $a_2^2/a_1^2$ yielding an accelerating universe. This can easily be understood from
figure \ref{lines}.
In general, a solution will only have transient acceleration. The
only exception is when ``Rome" lies within or on the ``arctic
circle", corresponding to $\Delta\leq-2$. Then, from the moment the
line crosses the ``arctic circle", there will be eternal acceleration
\cite{Townsend:2001ea} towards ``Rome". When $\Delta=-2$, there
will only be eternal acceleration when ``Rome" is approached from
the left. The possibilities of acceleration can be summarised as:
\begin{align}\nonumber
\bullet & \quad\Delta >-2 \, : & & {\rm A \ phase\ of\ transient\ acceleration\ is\
possible}\,,\\\nonumber \bullet & \quad\Delta =-2 \, : & & {\rm A\ phase\ of\ eternal\
acceleration\ is\ possible}\,,\\\nonumber \bullet & -3 \leq \Delta <-2 \, : & & {\rm Always\ a\
phase\ of\ eternal\ acceleration}\,.
\end{align}
The phase of eternal acceleration can also be understood from the power-law behaviour of the
``Rome" solution, i.e.~$a(\tau)\propto\tau^{1/(3+\Delta)}$. We have asymptotic acceleration when
$1/(3+\Delta)>1$, i.e.~$\Delta<-2$.
 In the limiting case $\Delta=-3$, corresponding to
``Rome" being on the North Pole, the interpolating solution will
asymptote to De Sitter.

\subsection{Equation of State}\label{eos}

In a cosmological setting, one often writes the matter part of the
equations in terms of a perfect fluid, which is described by its
pressure $p$ and energy density $\rho$. These two variables are
then assumed to be related via the equation of state:
\begin{align}
p=\kappa\,\rho\,.
\end{align}
As is well known in standard cosmology, $\kappa=0$ corresponds to the matter dominated era,
$\kappa=1/3$ to the radiation dominated era and $\kappa=-1$ to an era dominated by a pure
cosmological constant. Quintessence is a generalisation of the latter with $-1 \leq \kappa < -1/3$.

In our case, the matter is given by the two scalar fields, and thus $p$ and $\rho$ are given by the difference and sum of the kinetic terms and the potential, respectively:
\begin{align}
p=\tfrac{1}{2}\,(\dot{\varphi}^2+\dot{\phi}^2)-V\,,\quad
\rho=\tfrac{1}{2}\,(\dot{\varphi}^2+\dot{\phi}^2)+V\,.
\end{align}
Writing the above in terms of $x,y$ and $z$, we see that the
scalars describe a perfect fluid with an equation of state given
in terms of the parameter:
\begin{align}
\kappa=1-2z^2\,.
\end{align}
Hence, $\kappa$ varies from 1 on the equator to $-1$ on the North
Pole, and we need $\kappa<-1/3$ for quintessence. For the
interpolating solutions, which are given as curves on the sphere,
$\kappa$ will depend on time, but it will be constant for the
critical points with the following values \cite{Townsend:2001ea}:
\begin{alignat}{2}\nonumber
\bullet & \quad{\rm Equator} :\qquad &&\kappa=1 \,, \\
\bullet & \quad{\rm ``Rome"} :\qquad\quad
&&\kappa=1+\tfrac{2}{3}\,\Delta\,.
\end{alignat}

\subsection{One-scalar Truncations}\label{1scalar}

The analysis has so far been done for two scalars, and as such it also contains the truncation to a
system with one scalar with a potential, corresponding to $\phi=0$. Here we will summarise the
results of the previous sections in this truncation. On the sphere this yields $y=0$, and for the
solutions it corresponds to $a_2=b_2=0$.

\begin{figure}[h]
\centerline{\epsfig{file=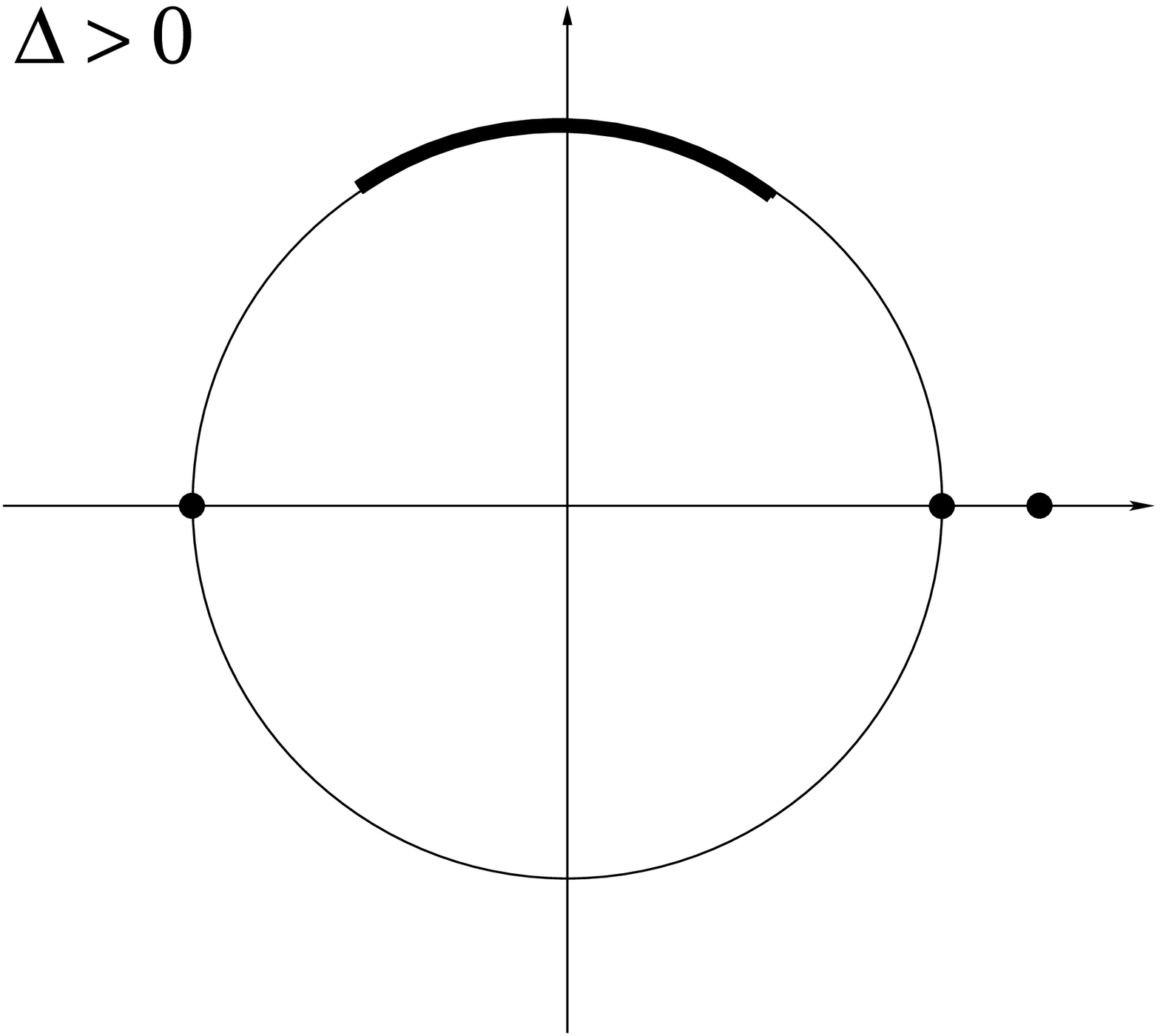,width=.33\textwidth}
\epsfig{file=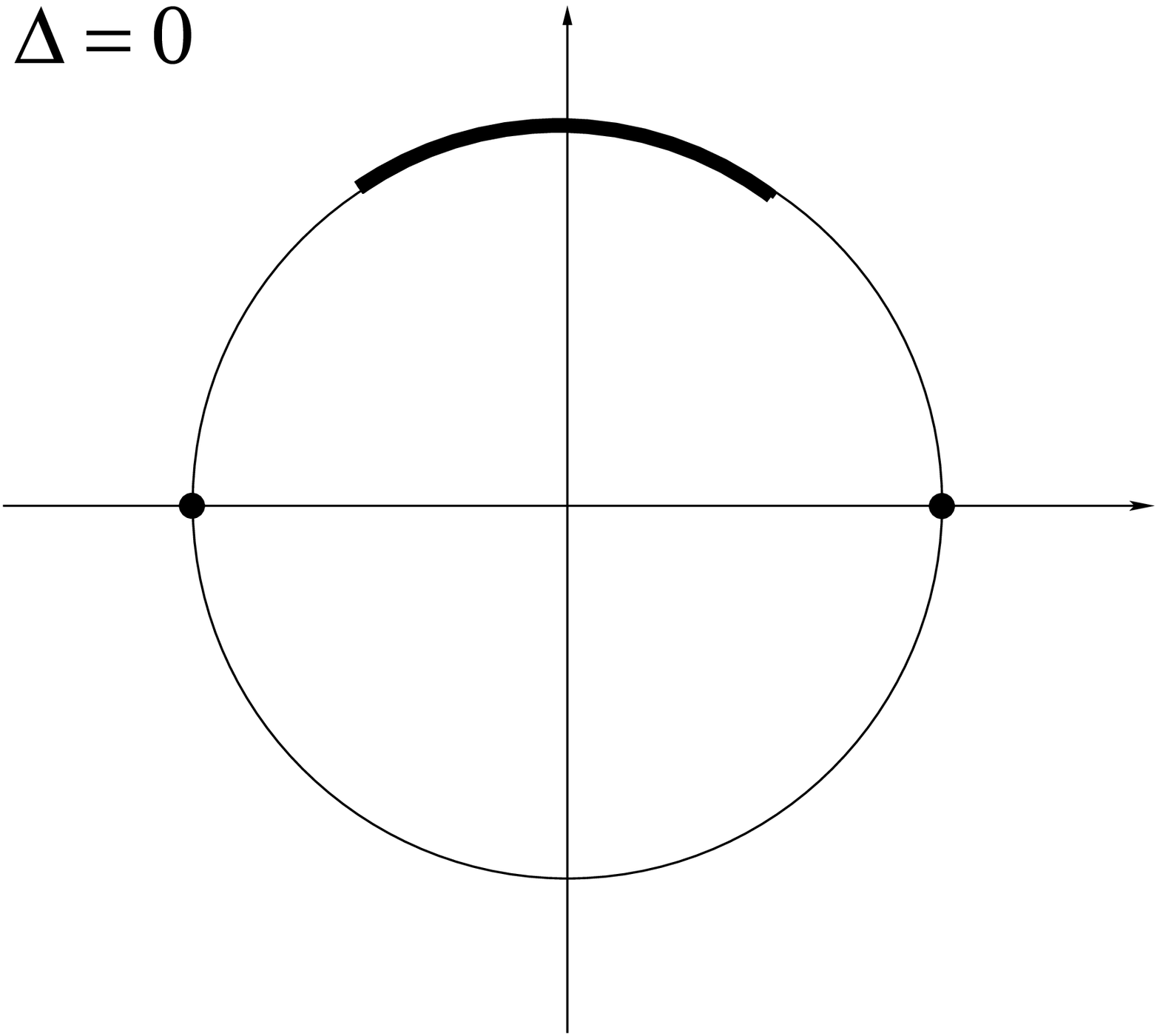,width=.33\textwidth} \epsfig{file=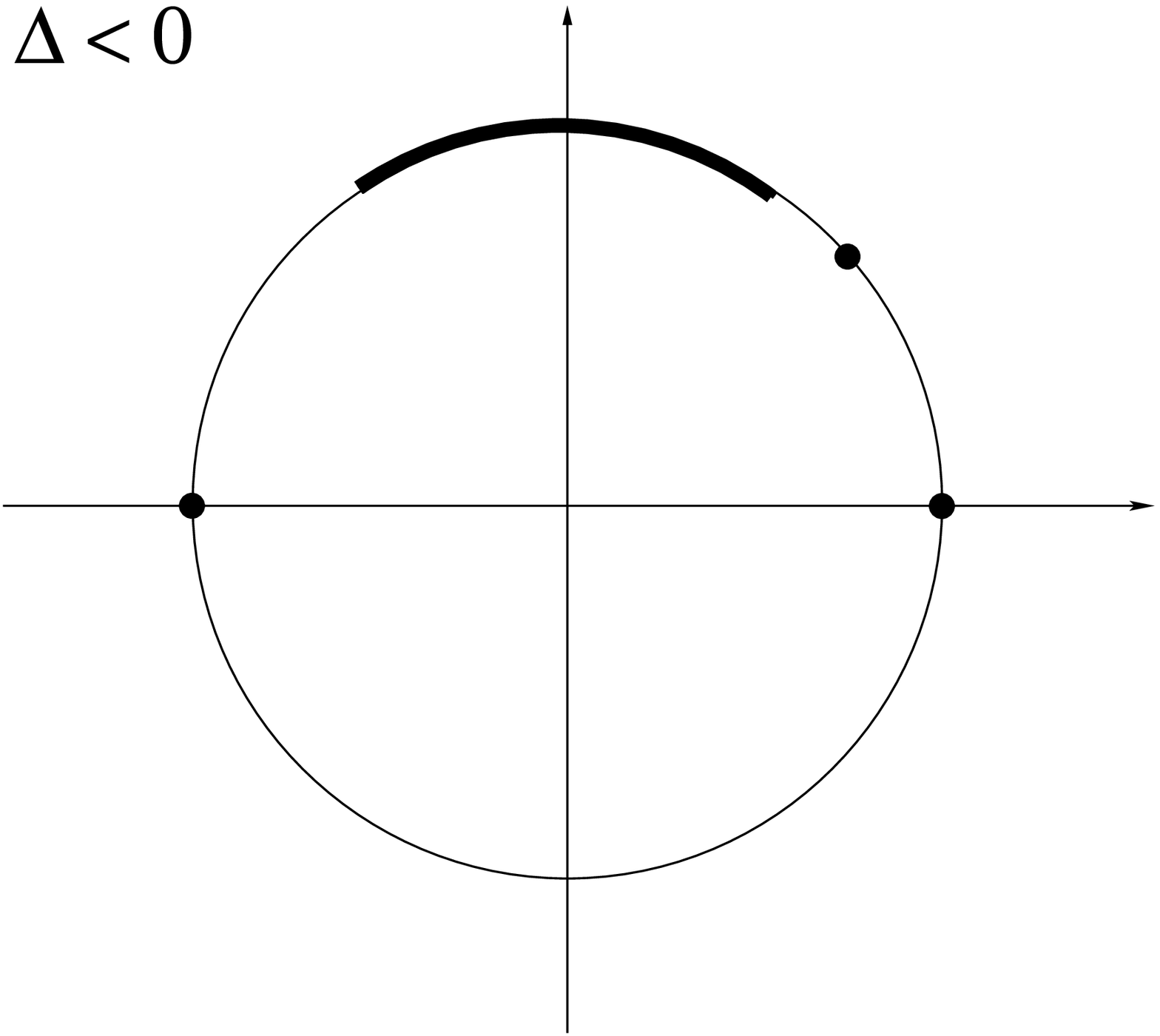,width=.33\textwidth}}
 \caption{\it The 2-dimensional $(x,z)$ space and the critical points for the one-scalar truncations.
The thick curve is the accelerating region. The two points on the
$x$-axis are the equatorial critical points. The third point is
``Rome". Note that in the middle illustration ``Rome" coincides
with the equatorial critical point $x=1$.}\label{1scphase}
\end{figure}

Since we only have one scalar, the Friedmann equation will define
a circle when written in terms of $x$ and $z$
\cite{Coley:1997nk,Heard:2002dr}:
\begin{align}
x^2+z^2=1\,.
\end{align}
The critical points are \cite{Halliwell:1987ja}:
\begin{alignat}{4}\nonumber
\bullet & \quad{\rm Equatorial} \, :& \qquad &z=0\,,\quad &&x^2&=&1\,,\\
\bullet & \quad{\rm ``Rome"} \, (-3\le\Delta<0) \, :& &z=\sqrt{-
\Delta / 3}\,,\quad &&x&=&\sqrt{1 + \Delta / 3}\,.
\end{alignat}
The full circle is shown in figure \ref{1scphase}, including the
critical points, and is just the vertical slice of the two-sphere
including the North Pole. The equator therefore becomes two
points, and the region bounded by the ``arctic circle" now becomes
the part of the circle\footnote{This is equivalent to the accelerating region of \cite{Cornalba:2003kd,Vieira:2003tx}.} corresponding to $x^2<1/3$.

The exact solutions can be obtained from the previous section by
setting $a_2=b_2=0$. For $\Delta\geq 0$ the solutions correspond
to the curves starting at $x=-1$ and ending at $x=1$, whereas for
$-3\le\Delta<0$ the curves start at either one of the equatorial
points and end at ``Rome". In all cases where the curve starts at
$x=-1$, the corresponding solution will give rise to acceleration. For this reason, interpolating solutions with $\Delta \geq 0$ will always give rise to a period of acceleration. This is in clear contrast to the two-scalar case, where it is possible to avoid acceleration (see figure \ref{lines}).
As for the 2-scalar case, if ``Rome" lies in the ``arctic" region,
the solution will be eternally accelerating from the moment it
enters this region.

One can also consider the truncation to zero scalars. However, from the scalar field equations, it
is seen that this is only consistent if $\Delta = -3$, and this corresponds to the De Sitter
solution with $V = \Lambda$.

\section{Higher-dimensional Origin}\label{M-theory}

In this section we will discuss the higher-dimensional origin of the 4D system (\ref{Lagrangian2})
by considering group manifold reductions. We will furthermore consider generalisations of the
single exponential potential to $n$-tuple exponential potentials with $n=2,3$ and $6$. In
particular, we will show that such potentials can be obtained by reduction of 7D gravity over the
3D group manifolds as classified by Bianchi. Furthermore, relations with S-branes will be
discussed. Finally, we will comment on the embedding of such potentials in M-theory.

We choose 3D group manifolds since these are well-studied and have
been classified. Two-dimensional group manifolds only include
$\mathbb{E}^2$ and $\mathbb{H}^2$ and are implicitly included in
our analysis. We anticipate the possibility of a similar analysis
for higher-dimensional group manifolds. Our approach is not to be
confused with Bianchi cosmology, in which everyday's 3D space is a
Bianchi group manifold; we employ the group manifold Ansatz to
reduce to 4D, in which we assume a flat FLRW space-time
\eqref{FRW}.

\subsection{Group Manifold Reductions} \label{gmr}

Group manifolds are homogeneous but in principle anisotropic
generalisations of maximally symmetric spaces, which are
homogeneous and isotropic. Therefore, the dimension of the
isometry group ${\cal G}^\text{isom}$, from which the group
manifold takes its name, in general equals the dimension of the
manifold. Only in special cases the manifold has more isometries
and can even be maximally symmetric. In this section we will take
the universal covering of the group manifolds (except for explicit
factors of $S^1$) and will not pay attention to global issues like
compactness; we will comment on this in section 4.

As was proven by Scherk and Schwarz \cite{Scherk:1979zr}, one can always dimensionally reduce a
theory containing gravity over a group manifold. This leads to a lower-dimensional theory with a
potential and a non-trivial gauge group $\cal G$. For reductions of pure gravity one finds $\cal G
= {\cal G}^\text{isom}$, but reductions of more general field contents can lead to bigger gauge
groups ${\cal G} \supseteq {\cal G}^\text{isom}$.

Turning to the reduction over a 3D group manifold, the structure
constants of the corresponding Lie algebra can be written as:
\begin{align}
  f_{mn}{}^p = \epsilon_{mnq} \mathsf{Q}^{pq} + 2 \delta_{[m}{}^p a_{n]} \,.
  \label{structure-constants}
\end{align}
By choosing an appropriate basis one can always take
$\mathsf{Q}^{pq} = \tfrac{1}{2} \text{diag}(q_1,q_2,q_3)$ with
$q_i = 0, \pm 1$ and $a_m = (a,0,0)$ (see
e.g.~\cite{Bergshoeff:2003ri}). The generators $T_m$ now satisfy
\begin{equation}
   [T_1 , T_2] = \tfrac{1}{2} q_3 T_3 -a T_2 \,, \qquad
   [T_2 , T_3] = \tfrac{1}{2} q_1 T_1 \,, \qquad
   [T_3 , T_1] = \tfrac{1}{2} q_2 T_2 +a T_3 \,.
\label{commutations}
\end{equation}
The Jacobi identity implies $q_1 a = 0$. This leads to 11 different algebras, given by the Bianchi
classification \cite{Bianchi, Ellis} in table \ref{3Dalgebras}. One distinguishes between algebras
of class A with $f_{mn}{}^n = 0$ and class B with $f_{mn}{}^n \neq 0$. For the class B algebras we
will use a notation that indicates that each class B algebra can be viewed as a deformation (with
deformation parameter $a$) of a class A algebra, see table \ref{3Dalgebras}.

\begin{table}[ht]
\begin{center}
\begin{tabular}{||c|c|c|c||c|c|c|c||}
\hline \rule[-3mm]{0mm}{8mm}
  Class A & a & $(q_1,q_2,q_3)$ & Algebra & Class B & a & $(q_1,q_2,q_3)$ & Algebra \\
\hline \hline \rule[-3mm]{0mm}{8mm}
  IX & 0 & $(1,1,1)$ & $so(3)$ & & & & \\
\hline \rule[-3mm]{0mm}{8mm}
  VIII & 0 & $(1,-1,1)$ & $so(2,1)$ & & & & \\
\hline \rule[-3mm]{0mm}{8mm}
  VII$_0$ & 0 & $(0,1,1)$ & $iso(2)$ & VII$_a$ & a & $(0,1,1)$ & $iso(2)_{a}$ \\
\hline \rule[-3mm]{0mm}{8mm}
  VI$_0$ & 0 & $(0,-1,1)$ & $iso(1,1)$ & III & 1/2 & $(0,-1,1)$ & $iso(1,1)_{a=1/2}$ \\
   & & & & VI$_a$ & a & $(0,-1,1)$ & $iso(1,1)_{a}$ \\
\hline \rule[-3mm]{0mm}{8mm}
 II & 0 & $(0,0,1)$ & $heisenberg$ & IV & 1 & $(0,0,1)$ & $heisenberg_{a=1}$ \\
\hline \rule[-3mm]{0mm}{8mm}
 I & 0 & $(0,0,0)$ & $u(1)^3$ & V & 1 & $(0,0,0)$ & $u(1)^3_{a=1}$ \\
\hline
\end{tabular}
\caption{\it The Bianchi classification of the different three--dimensional Lie algebras in terms
of components of their structure constants.}
\end{center}
\label{3Dalgebras}
\end{table}

Our starting point is 7D pure gravity $\hat{g}_{\hat{\mu}
\hat{\nu}}$ with the Einstein-Hilbert term as Lagrangian. For
dimensional reduction we split up indices in $\hat{\mu} = (\mu,m)$
where $x^\mu$ is the 4D space-time while $z^m$ corresponds to the
3D internal space. The reduction Ansatz from 7D to 4D reads:
 \begin{align}
   \hat{ds}^2 = e^{-3 \varphi / \sqrt{15}} g_{\mu \nu} dx^\mu dx^\nu
    + e^{2 \varphi / \sqrt{15}} {\cal M}_{pq} U^p{}_m U^q{}_n dz^m dz^n \,,
 \label{Ansatz}
 \end{align}
where we have truncated away the three Kaluza-Klein vectors. The resulting 4D field content is $\{
g_{\mu \nu}, {\cal M}_{mn}, \varphi \}$, where ${\cal M}_{mn}$ parametrises the coset
$SL(3,\mathbb{R}) / SO(3)$ and contains two dilatons and three axions; see
\cite{AlonsoAlberca:2003jq, Bergshoeff:2003ri} for an explicit representation in terms of the five
scalars. In all cases but one we will consider the following truncated parametrisation of the
scalar coset:
 \begin{align}
   {\cal M}_{mn} = \text{diag} (e^{-2 \sigma/\sqrt{3}},
  e^{-\phi+\sigma/\sqrt{3}},e^{\phi+\sigma/\sqrt{3}})\,,
 \label{truncation}
 \end{align}
where we have set the axions equal to zero\footnote{This will be a consistent truncation if $a \,
(q_2 e^{- \phi} - q_3 e^{+\phi}) = 0$, which is satisfied for all class A types ($a=0$), class B
type V ($q_2=q_3=0$) and type VII$_a$ (for $\phi = 0$). \label{axiontruncation}}. The matrix
$U^m{}_n (z^p)$ contains the only dependence on the internal manifold \cite{Bergshoeff:2003ri}:
\begin{align}
  U^m{}_n = \left(
  \begin{array}{ccc}
    1 & 0 &-s_{1,3,2} \\
    0 & e^{a z^1}\,c_{2,3,1} & e^{a z^1}\,c_{1,3,2} \, s_{2,3,1} \\
    0 & -e^{a z^1}\,s_{3,2,1} & e^{a z^1}\,c_{1,3,2} \, c_{2,3,1}
  \end{array} \right) \,,
\label{explicitU}
\end{align}
where we have used the following abbreviations:
\begin{align}
  c_{m,n,p} \equiv \cos(\tfrac{1}{2} \sqrt{q_m}\sqrt{q_n} \, z^p) \,, \qquad
  s_{m,n,p} \equiv \sqrt{q_m} \sin(\tfrac{1}{2}\sqrt{q_m}\sqrt{q_n} \, z^p) /\sqrt{q_n} \, ,
\end{align}
and it is understood that the structure constants satisfy the
Jacobi identity, amounting to $q_1 a =0$. The $U^m{}_n$'s
parametrise the different group manifolds and consist of the
components of the Maurer-Cartan 1-forms. After reduction the
$U^m{}_n$'s only appear in the $z^m$-independent combination:
\begin{align}
  f_{mn}{}^{p} = -2(U^{-1})^r{}_{m} (U^{-1})^s{}_n\, \partial_{[r} U^p{}_{s]} \,,
 \label{MC}
\end{align}
where $f_{mn}{}^{p}$ are the structure constants \eqref{structure-constants}.

Before diving into the dimensional reduction we would like to address the issue of consistent
truncations. A truncation is considered to be consistent if any solution to the truncated field
equations leads to a solution of the full field equations \cite{AlonsoAlberca:2003jq,
Cvetic:2003jy}. All truncations we consider in this section are consistent in this sense. \\

The reduction over a class A group manifold results in the
following Lagrangian: \cite{AlonsoAlberca:2003jq}
 \begin{align}
  {\cal L}_\text{class A} = \sqrt{-g} \, [ R -\tfrac{1}{2} (\partial \vec{\phi})^2 + \tfrac{1}{2} \, e^{- \sqrt{5} \varphi / \sqrt{3}}\, \{ \left[ {\rm Tr}({\cal M}{\mathsf Q}) \right]^2
   - 2 {\rm Tr}({\cal M}{\mathsf Q}{\cal M}{\mathsf Q})\}] \,,
 \label{potentialQ}
 \end{align}
with $\vec{\phi} = (\varphi, \sigma, \phi)$. The scalar potential is always of multiple exponential
form in the truncation \eqref{truncation}. In table \ref{truncations} we list the different
possibilities to truncate the general scalar potential of class A. We will elaborate on these
truncations in the next subsections. Note that the group manifolds generically gain isometries in
these truncations; we also give the resulting manifold.

\begin{table}[ht]
\begin{center}
\begin{tabular}{||c||c|c||c|c||}
\hline \rule[-3mm]{0mm}{8mm}
  Bianchi & Truncation \eqref{truncation} & Potential & \# Isometries & Manifold \\
\hline \rule[-3mm]{0mm}{8mm}
  IX & - & 6-tuple & $3$ & $so(3)$ \\
   & $\phi=0$ & double & $4$ & $S^2 \Join S^1$ \\[0.6ex]
   & $\phi=\sigma=0$ & single & $6$ & $S^3$ \\
\hline \rule[-3mm]{0mm}{8mm}
  VIII & - & 6-tuple & $3$ & $so(2,1)$ \\
   & $\phi=0$ & double & $4$ & $\mathbb{H}^2 \Join S^1$ \\
\hline \rule[-3mm]{0mm}{8mm}
  VII$_0$ & - & triple & $3$ & $iso(2)$ \\
   & $\phi =0$ & - & $6$ & $\mathbb{E}^3$ \\
\hline \rule[-3mm]{0mm}{8mm}
  VI$_0$ & - & triple & $3$ & $iso(1,1)$ \\
   & $\phi =0$ & single & $3$ & $iso(1,1)$ \\
\hline \rule[-3mm]{0mm}{8mm}
  II & - & single & $4$ & $\mathbb{E}^2 \Join S^1$\\
\hline \rule[-3mm]{0mm}{8mm}
  I & - & - & $6$ & $\mathbb{E}^3$ \\
\hline
\end{tabular}
\caption{\label{truncations}\it The different truncations of the
class A group manifolds leading to multiple exponential
potentials. The table also gives the corresponding internal
manifold and its number of isometries. Here $\times$ and $\Join$
denote the direct and twisted product, respectively (see also
section \ref{double-section}). The coset manifolds $S^n$ and
$\mathbb{H}^n$ are understood to have the maximally symmetric
metric. The cases with $4$ and $6$ isometries of type IX are
called the squashed and round $S^3$, respectively.}
\end{center}
\end{table}

The reduction over a class B group manifold gives rise to a less familiar situation: one can not
even write down a Lagrangian, but only field equations that can not be derived as Euler-Lagrange
equations \cite{AlonsoAlberca:2003jq}. From the massive deformations of the Einstein equations one
can find a generalised class B-version of the potential of \eqref{potentialQ}, including
$a$-dependent terms, but the deformations of the scalar field equations are not the derivatives of
this potential. However, we will encounter truncations for which the construction of a Lagrangian
is possible. In these cases the truncation ensures that the massive deformations of the scalar
field equations indeed are the derivatives of the potential derived from the Einstein equations.

\subsection{Single Exponential Potentials}
 \label{truncation-single}

We first consider the possible truncations of the group manifold reductions to the system
\eqref{LagrangianN} with a single exponential potential:
 \begin{align}
  {\cal L}_\text{single} = \sqrt{-g} \, [ R -\tfrac{1}{2} (\partial \vec{\phi})^2 - \Lambda \, e^{\vec{\alpha} \cdot \vec{\phi}} ]\,.
 \label{single}
 \end{align}
As explained in section \ref{Multi-Scalar} one can always choose a basis such that only one scalar
appears in the potential. To avoid the need to do so we define the generalised parameter $\Delta =
(\vec{\alpha} \cdot \vec{\alpha} - 3)_\pm$, which will be invariant under changes of basis. The
subscript $\pm$ denotes the sign of $\Lambda$.

The potential of class A is in general more complicated than the
single exponential potential that we have considered in section 2.
However, there are three different possibilities to truncate the
potential given in \eqref{potentialQ} to \eqref{single}. These are
the type II, type VI$_0$ and type IX cases, as given in table
\ref{single-table}. To be concrete we give the example of type II,
whose Lagrangian with the truncation \eqref{truncation} reads:
\begin{align}
  {\cal L} = \sqrt{-g} \, [R - \tfrac{1}{2} (\partial \varphi)^2 - \tfrac{1}{2} (\partial \phi)^2
  - \tfrac{1}{2} (\partial \sigma)^2 - \tfrac{1}{8} e^{- \sqrt{5} \varphi / \sqrt{3} + 2 \sigma \sqrt{3} + 2 \phi}]
  \,.
 \label{typeII}
 \end{align}
Similar formulae can be derived from \eqref{potentialQ} in the other cases. The only possibility to
obtain a negative potential of single exponential form occurs in the type IX potential. This
corresponds to the gauging of $SO(3)$, which is the only compact gauge group of the Bianchi
classification. Note that the truncation $\sigma = \phi = 0$, leading to a single exponential
potential, corresponds to the case of maximal isometry enhancement such that the group manifold
becomes the round $S^3$, see table \ref{truncations}.

\begin{table}[ht]
\begin{center}
\begin{tabular}{||c|c|c|c||c||}
\hline \rule[-3mm]{0mm}{8mm}
  Bianchi & Class& Truncation & $\Delta$ & Manifold \\
\hline \hline \rule[-3mm]{0mm}{8mm}
  II & A&\eqref{truncation} & $4_+$ & $\mathbb{E}^2 \Join S^1$ \\
\hline \rule[-3mm]{0mm}{8mm}
VI$_0$&A&\eqref{truncation}, $\phi=0$&$0_+$ & $iso(1,1)$\\
\hline \rule[-3mm]{0mm}{8mm}
  III & B&\eqref{typeIIItrunc} & $-1_+$ & $\mathbb{H}^2 \times S^1$ \\
\hline \rule[-3mm]{0mm}{8mm}
  V & B&\eqref{truncation}, $\sigma = \phi = 0$ & $-4/3_+$ & $\mathbb{H}^3$ \\
\hline \rule[-3mm]{0mm}{8mm}
  VII$_a$ & B&\eqref{truncation}, $\sigma = \phi = 0$ & $-4/3_+$ & $\mathbb{H}^3$ \\
\hline \hline \rule[-3mm]{0mm}{8mm}
  IX & A&\eqref{truncation}, $\sigma = \phi = 0$ & $-4/3_-$ & $S^3$ \\
\hline
\end{tabular}
\caption{\it The truncations to a single exponential potential of positive or negative sign and
the reduction manifold corresponding to these truncations.} \label{single-table}
\end{center}
\end{table}

The full set of field equations of class B gaugings cannot be derived from a Lagrangian. However,
for specific truncations this is possible. We know of three such cases leading to a Lagrangian with
a single exponential potential, see table \ref{single-table}. The truncation of type III is not of
the form \eqref{truncation} but given by\footnote{
 Note that the type III case does not satisfy the constraint of footnote
\ref{axiontruncation}. The off-diagonal components of $\cal M$ (corresponding
 to non-zero axions) are a consequence of our basis choice for the structure constants. A simple
 $SO(2)$-rotation brings $\cal M$ to diagonal form but introduces off-diagonal components in
 $\mathsf{Q}$.}:
  \begin{align}
  {\cal M} = \left( \begin{array}{ccc}
  e^{-\sigma/\sqrt{3}} & 0 & 0 \\
  0 & e^{\sigma / 2 \sqrt{3}} \text{cosh}(\tfrac{1}{2} \sqrt{3} \sigma) & -e^{\sigma / 2 \sqrt{3}} \text{sinh}(\tfrac{1}{2} \sqrt{3} \sigma) \\
  0 & -e^{\sigma / 2 \sqrt{3}} \text{sinh}(\tfrac{1}{2} \sqrt{3} \sigma) & e^{\sigma / 2 \sqrt{3}} \text{cosh}(\tfrac{1}{2} \sqrt{3} \sigma)
  \end{array} \right) \,.
  \label{typeIIItrunc}
  \end{align}
In this truncation, the type III manifold becomes $ \mathbb{H}^2 \times S^1$. Similarly, the
truncations of type V and type VII$_a$ correspond to the manifold $\mathbb{H}^3$. Thus, in all
three cases in which class B allows for an action, the group manifold (partly) reduces to a
hyperbolic manifold. We furthermore see that in all cases a single exponential potential
requires a group manifold with maximum isometry enhancement. \\

In summary, we find that the different group manifold reductions give rise to examples of the three
classes of potentials of the form in \eqref{LagrangianN} with $\Delta$ positive, zero or negative.
In particular we find the following possibilities:
 \begin{align}
  \Delta = -\tfrac{4}{3} , -1 , 0, 4,
 \end{align}
for positive potentials.

A very natural question concerns the uplift of the solutions in section 2 to 7D. By construction,
all solutions with any of the $\Delta$-values with subscript $+$ listed in table \ref{single-table}
can be uplifted to a purely gravitational solution in 7D. Of these, the interpolating solution with
$\Delta = -4/3$ uplifts to the fluxless limit of the S2-brane solution in 7D \cite{Ohta:2003pu}. We
will comment on the uplift of the interpolating solutions with $\Delta = -1,0,4$ in the next two
subsections.

In addition, we have checked that the critical point solutions with $\Delta = -4/3,-1$ uplift to a
7D (locally) Minkowskian space-time. This is similar to the results of \cite{Bergshoeff:2003ri},
where an 8D cosmological power-law solution uplifted (via $\mathbb{H}^3$) to 11D (locally)
Minkowskian space-time. In fact, the $\Delta = -4/3$ case can be understood as a $T^4$ reduction of
this, see also section 3.5. The $\Delta = -1$ case will have an analogous power-law solution in 8D
and 9D, which will also uplift (via $\mathbb{H}^2 \times S^1$ and $\mathbb{H}^2$, respectively) to
11D (locally) Minkowskian space-time.

The above results can be understood in terms of the Milne parametrisation of flat space-time. In this coordinate system the spatial part of Minkowski space-time is a hyperbolic space. Adding a three-dimensional flat space to an $(n+1)$-dimensional Milne space-time, one can reduce over the hyperbolic part to a four-dimensional cosmological solution \cite{Russo:2003ky}. In fact, this is similar to the reduction of $D=11$ Minkowski space-time to the $\Delta=-4/3_-$ domain wall solution in eight dimensions \cite{Boonstra:1998mp}. Here flat space-time is written in polar coordinates allowing for a reduction over an $S^3$ instead of a hyperbolic space. This is the domain wall analogue of the cosmological ``Rome" solution.

\subsection{Double Exponential Potentials and S-branes} \label{double-section}

In this subsection we consider the truncation to double exponential potentials. Furthermore we
point out the relation between certain group manifold reductions and another method of dimensional
reduction, namely over a maximally symmetric space with flux \cite{Bremer:1998zp} and
S-branes. \\

We first consider the generalisation of the 4D system
\eqref{Lagrangian2} to double exponential potentials. The
generalised Lagrangian reads:
 \begin{align}
  {\cal L}_\text{multi} = \sqrt{-g} \, [ R -\tfrac{1}{2} (\partial \vec{\phi})^2 - \sum_{i=1}^n \Lambda_i \, e^{\vec{\alpha}_i \cdot \vec{\phi}}
  ]\,,
 \label{multi}
 \end{align}
with $n=2$. Following section \ref{Multi-Scalar} we define
$\Delta_i = (\vec{\alpha}_i \cdot \vec{\alpha}_i - 3)_\pm$, where
the subscript $\pm$ denotes the sign of $\Lambda_i$. The general
class A potential \eqref{potentialQ} gives rise to two potentials
of this form, see table \ref{double-table}. The coefficients
$\Lambda_i$ can be taken to have the same absolute value. Due to
the opposite signs of the exponentials in the type IX case, one of
the scalars has a minimum and indeed this case can be truncated to
a single exponential potential (with a negative sign, see table
\ref{single-table}). We know of no class B truncations to a double
exponential potential.

\begin{table}[ht]
\begin{center}
\begin{tabular}{||c|c|c|c|c|c||c||}
\hline \rule[-3mm]{0mm}{8mm}
  Bianchi & Class&Truncation & $\Delta_1$ & $\Delta_2$ & $\alpha_1 \cdot \alpha_2$ & Manifold \\
\hline \hline \rule[-3mm]{0mm}{8mm}
  VIII & A&\eqref{truncation}, $\phi = 0$ & $4_+$ & $-1_+$ & $3$ & $\mathbb{H}^2 \Join S^1$ \\
\hline \rule[-3mm]{0mm}{8mm}
  IX & A&\eqref{truncation}, $\phi = 0$ & $4_+$ & $-1_-$ & $3$ & $S^2 \Join S^1$ \\
\hline
\end{tabular}
\caption{\it The different truncations to double exponential potentials and the reduction manifold
corresponding to these truncations.} \label{double-table}
\end{center}
\end{table}

The reduction schemes leading to these double potentials can also
be seen from another point of view. Due to the independence of the
group manifold Ansatz \eqref{explicitU} of $z^3$, one can first
perform a trivial reduction to 6D. This results in the Lagrangian:
 \begin{align}
  {\cal L}_6 = \sqrt{-\hat{g}} \, [ \hat{R} - \tfrac{1}{2} (\partial \hat{\sigma})^2
  - \tfrac{1}{4} e^{- \sqrt{5} \hat{\sigma} / \sqrt{2}} \hat{F}_2{}^2 ] \,,
  \label{6D}
 \end{align}
where $\hat{F}_2$ is the field strength of the Kaluza-Klein vector. All Bianchi group manifold
reductions thus correspond to a non-trivial reduction of this 6D system. A subset of those will
correspond to reductions over a maximally symmetric 2D space with flux, having the following
reduction Ansatz:
 \begin{align}
  \hat{ds}_6^2 = e^{-\varphi/\sqrt{2}} ds_4^2 + e^{\varphi/\sqrt{2}} d \Sigma_k^2 \,, \qquad
  \hat{F}_2 = \sqrt{2} \, f \, \text{vol}(\Sigma_k) \,, \qquad \hat{\sigma} = \sigma \,,
 \end{align}
where the parameter $f$ is the flux through the internal 2D manifold, which has constant curvature
$k = 0, \pm 1$: we take $R_{ab} (\Sigma_k) = \tfrac{1}{2} \, k \, g_{ab}$. This results in a
lower-dimensional theory with a double exponential potential \cite{Bremer:1998zp, Roy:2003nd}:
 \begin{align}
  {\cal L}_{4} = \sqrt{-g} \, [ R - \tfrac{1}{2} (\partial \sigma)^2 - \tfrac{1}{2} (\partial \varphi)^2
  - f^2 \, e^{-\sqrt{5} \sigma/\sqrt{2} - 3 \varphi / \sqrt{2} } + k \, e^{-\sqrt{2} \varphi} ] \,.
 \label{Sbranepotentialfrom6D}
 \end{align}
For $f \neq 0$ and $k \neq 0$ this double exponential potential has the same characteristics as
given in table \ref{double-table} with the sign of $\Lambda_2$ given by the sign of $-k$. For
either of the potential terms vanishing one finds one of the cases of table \ref{single-table}.

By comparison with tables \ref{single-table} and \ref{double-table} one can relate the group
manifold reductions to flux reductions over maximally symmetric 2D spaces, i.e.~$S^2, \mathbb{E}^2$
or $\mathbb{H}^2$. Perhaps the most striking relation is the equivalence between the type IX group
manifold with the truncations \eqref{truncation} and $\phi = 0$ and the flux reduction with $f \neq
0$ and $k = +1$ \cite{Boonstra:1998mp}. This corresponds to the three-sphere as the Hopf fibration
over $S^2$, denoted as $S^2 \Join S^1$ in table \ref{truncations}. Due to the truncation $\phi =
0$, one gains an extra isometry and therefore this reduction is over the squashed three-sphere, see
also table \ref{truncations}. The further truncation to $\sigma = 0$ brings one to the round $S^3$
and single exponential potential of section \ref{truncation-single}.

Without the flux the 3D manifold becomes the direct product $S^2 \times S^1$, which gives rise to a
single exponential potential with $\Delta = -1_-$ in table \ref{single-table}. However, this is not
a group manifold and therefore reduction over it does not necessarily preserve
supersymmetry\footnote{Supersymmetry would require $f \sim k$ but we have $f=0$ and $k=+1$.}; the
resulting potential will generically not be embeddable within a gauged supergravity.

Similarly, the type VIII group manifold with the truncations \eqref{truncation} and $\phi = 0$ is
equivalent to the flux reduction with $f \neq 0$ and $k = -1$. Thus the $so(2,1)$ group manifold
can be seen as a fibration over $\mathbb{H}^2$, denoted as $\mathbb{H}^2 \Join S^1$. Its fluxless
limit yields the direct product $\mathbb{H}^2 \times S^1$, which is a type III group manifold and
indeed yields a single exponential potential with $\Delta = -1_+$ in table \ref{single-table}.

The remaining case is the type II group manifold, which can be seen as a fibration over
$\mathbb{E}^2$, denoted as $\mathbb{E}^2 \Join S^1$. This yields a single exponential potential
with $\Delta = 4_+$. Taking the zero flux limit leads to the direct product $\mathbb{E}^2 \times
S^1$, which is a group manifold of type I, leading to a vanishing potential.

\vspace{5mm}

Having shown that the above double exponential potentials have a higher-dimensional origin in six
or seven dimensions, it is interesting to uplift any cosmological solution. Solutions to the single
and double exponential cases uplift to various S-branes in 6D \cite{Lu:1997jk, Lu:1996er,
Ivashchuk:1999xp, Ohta:2003pu, Roy:2003nd, Ohta:2003ie}. In particular, the interpolating solution
with $\Delta = -1$ uplifts to the fluxless limit of the hyperbolic S2-brane in 6D.
Again, the critical point solution with $\Delta = -1$ uplifts to a (locally) Minkowskian
space-time in 6D \cite{Russo:2003ky}.

{From} the above dictionary, one can uplift all these solutions one step further: to a purely
gravitational solution in 7D. It would be very interesting to identify which geometry the
interpolating solutions correspond to.

\subsection{Triple Exponential Potentials and Exotic S-branes}\label{triple}

We now turn to the truncations to triple exponential potentials. These will turn out to be related
to reductions over a circle with different $SL(2,\mathbb{R})$ twists.

The truncations we are interested in have the Lagrangian \eqref{multi} with $n=3$. The class A
potential \eqref{potentialQ} gives rise to two different triple exponential potentials, which are
given in table \ref{triple-table} and explicitly read (the class A Lagrangian \eqref{potentialQ}
with $q_1 = 0$):
 \begin{align}
  {\cal L} = \sqrt{-g} \, [ R - \tfrac{1}{2} (\partial \vec{\phi})^2
  - \tfrac{1}{8} e^{- \sqrt{3} \varphi} (q_2 e^{-\phi} - q_3 e^{\phi})^2 ] \,.
  \label{potential-triple}
 \end{align}
Both triple exponential potentials allow for the truncation $\phi = 0$. For the type VI$_0$ case
with $q_2 = - q_3$ this results in a single exponential potential with $\Delta = 0_+$ (as given in
table \ref{single-table}). For the type VII$_0$ case with $q_2 = q_3$ the terms cancel and one is
left without a potential. In this truncation, the type VII$_0$ group manifold reduces to the type I
manifold, in complete analogy to the degeneration of type VII$_a$ to type V in the truncation $\phi
= \sigma = 0$. As in the double exponential case, we know of no class B truncations to a triple
exponential potential.

\begin{table}[ht]
\begin{center}
\begin{tabular}{||c|c|c|c|c|c|c|c|c||c||}
\hline \rule[-3mm]{0mm}{8mm}
  Bianchi & Class&Truncation & $\Delta_1$ & $\Delta_2$ & $\Delta_3$
  & $\vec{\alpha}_1 \cdot \vec{\alpha}_2$ & $\vec{\alpha}_1 \cdot \vec{\alpha}_3$ & $\vec{\alpha}_2 \cdot \vec{\alpha}_3$ & Manifold \\
\hline \hline \rule[-3mm]{0mm}{8mm}
  VI$_0$ & A&\eqref{truncation} & $4_+$ & $0_+$ & $4_+$ & 3 & -1 & 3 & $iso(1,1)$ \\
\hline \rule[-3mm]{0mm}{8mm}
  VII$_0$& A& \eqref{truncation} & $4_+$ & $0_-$ & $4_+$ & 3 & -1 & 3 & $iso(2)$ \\
\hline
\end{tabular}
\caption{\it The different truncations to triple exponential potentials and the reduction manifold
corresponding to these truncations.} \label{triple-table}
\end{center}
\end{table}

Similar to the case of double exponential potentials, the above
triple exponential potentials can be obtained from another point
of view, which in this case is five-dimensional. The group
manifold Ansatz \eqref{explicitU} with $q_1 =0$ is independent of
$z^3$ and $z^2$. Therefore, one can always reduce to 5D where the
intermediate Lagrangian reads:
 \begin{align}
  {\cal L}_{5}
  & = \sqrt{-\hat{g}} \, [ \hat{R} - \tfrac{1}{2} (\partial \hat{\sigma})^2 - \tfrac{1}{2} (\partial \hat{\phi})^2
    - \tfrac{1}{2} e^{2 \hat{\phi}} (\partial \hat{\chi})^2 ] \,, \notag \\
  & = \sqrt{-\hat{g}} \, [ \hat{R} - \tfrac{1}{2} (\partial \hat{\sigma})^2
    + \tfrac{1}{4} \, \text{Tr}(\partial \hat{\cal K} \partial \hat{\cal K}^{-1}) ] \,,
 \label{5D}
 \end{align}
where we have truncated away the two Kaluza-Klein vectors and
defined:
 \begin{align}
  \hat{\cal K} = e^{\hat{\phi}} \left(
  \begin{array}{cc} e^{-2 \hat{\phi}} + \hat{\chi}^2 & \hat{\chi} \\ \hat{\chi} & 1 \end{array}
  \right) \,.
 \end{align}
Thus every Bianchi group manifold reduction with $q_1 = 0$ corresponds to a reduction of this 5D
system.

The 5D system \eqref{5D} has a global symmetry $\hat{\cal K} \rightarrow \Omega \hat{\cal K}
\Omega^T$ with $\Omega \in SL(2,\mathbb{R})$. In addition the 5D field equations are invariant
under the $\mathbb{R}^+$ transformation $\hat{g}_{\hat{\mu} \hat{\nu}} \rightarrow \lambda
\hat{g}_{\hat{\mu} \hat{\nu}}$. Note that this transformation scales the Lagrangian but leaves the
equations of motion invariant.

Every generator of a global symmetry can be employed for a reduction over a circle with a twist
\cite{Scherk:1979ta}. Restricting this generator to $SL(2,\mathbb{R})$ gives rise to three distinct
lower-dimensional theories, with different gauge groups and different potentials \cite{Hull:2002wg,
Bergshoeff:2002mb}. We will focus on these three cases in this subsection. If one also includes the
$\mathbb{R}^+$ generator one obtains a lower-dimensional theory with only field equations
\cite{Bergshoeff:2002nv}. This corresponds to a reduction over a class B group manifold. However,
we will not consider these here.

The reduction Ansatz with an $SL(2,\mathbb{R})$ twist reads
\cite{Meessen:1998qm}:
 \begin{align}
  \hat{ds}_5^2 = e^{-\varphi/\sqrt{3}} ds_4^2 + e^{2\varphi/\sqrt{3}} (dz^1)^2 \,, \quad
  \hat{\sigma} = \sigma \,, \quad
  \hat{\cal K} = \Omega {\cal K}
  \Omega^T \,, \quad
  \Omega = \text{exp} \left(
  \begin{array}{cc} 0 & q_2 z^1 \\ q_3 z^1 & 0 \end{array}
  \right) \,,
 \end{align}
with ${\cal K} = \text{diag}(e^{-\phi}, e^{\phi})$. Here the two parameters $q_2$ and $q_3$
distinguish between the three different subgroups of $SL(2,\mathbb{R})$:
 \begin{align}
  (q_2,q_3) =
  \begin{cases}
   (0,1): \ \ \Omega \in \mathbb{R} \,, \\
   (1,1): \ \ \Omega \in SO(2) \,, \\
   (-1,1): \Omega \in SO(1,1)^+ \,.
  \end{cases}
 \end{align}
Indeed, this reduction results in \eqref{potential-triple} for the Bianchi types II, VI$_0$ and
VII$_0$ with appropriate values of $q_2$ and $q_3$. Thus, we have shown that reduction over these
specific Bianchi types corresponds to a twisted reduction from 5D with one of the three subgroups
of $SL(2,\mathbb{R})$. The use of the $\mathbb{R}$ subgroup corresponds to the flux
compactification with flux $d \chi$.

We would like to comment on the 5D point of view of the generation of the single exponential
potential $\Delta = 0_+$. The group manifold Ansatz in the truncations \eqref{truncation} and $\phi
= 0$ boils down to a 5D reduction Ansatz with vanishing axion and a dilaton which only takes values
in the internal $z^1$ space. This can be seen as a flux compactification where the dilaton
$\hat{\phi}$ takes the role of the gauge potential, leading to $\Delta = 0_+$.

\vspace{5mm}

Let us now discuss the uplift of solutions of the triple exponential potentials of types VI$_0$ and
VII$_0$ and the single exponential potential of type II to five and seven dimensions. The type II
solution of section 2 will uplift to an S2-brane in 5D, which has not been given in the
literature\footnote{The standard formula for $Sp$-brane solutions in $D$ dimensions breaks down for
$p=D-3$.}. Solutions to the triple exponential case will uplift to exotic S2-branes in 5D, but such
solutions have not yet been constructed. The fact that $D=5$ is not relevant in this construction:
the axion-dilaton-gravity system will allow three such S$(D-3)$-solutions in arbitrary $D$.

Clearly, one can uplift all these solutions to purely gravitational solutions in 7D and it would be
very interesting to identify which geometry the different solutions correspond to.

\subsection{Embedding in M-theory}
 \label{gsg}

As discussed, the reduction of 7D gravity over a group manifold gives rise to a gauge group and a
scalar potential in 4D. The 7D gravity can be trivially embedded in ungauged 7D maximal
supergravity, which in itself arises from a trivial $T^4$ reduction of 11D supergravity, the
low-energy limit of M-theory. Therefore, the 4D systems under consideration have an origin in
M-theory. Moreover, since a group manifold reduction does not break supersymmetry\footnote{
  At least when reducing over the universal covering space of group manifolds. When considering
  reductions over compactified versions, supersymmetry is not necessarily preserved; see section 4.},
the 4D systems can be embedded in a gauged maximal supergravity. In addition, the generalisation to
more general field contents can lead to bigger gauge groups ${\cal G} \supseteq {\cal
G}^\text{isom}$. The following example illustrates these points.

Reduction of 7D ungauged maximal supergravity over a class A group manifold leads to 4D gauged
supergravities. The gauge group is $CSO(p,q,8-p-q)$ for $p$ positive entries and $q$ negative
entries in the mass matrix\footnote{
 See the appendix in \cite{AlonsoAlberca:2003jq} for the relation between the Bianchi classification
 and $CSO(p,q,r)$ groups \cite{Hull:1985rt}.}
$Q^{mn}$ and therefore $p+q \leq 3$. The same theory can be obtained by a reduction over $T^4$ of
the 8D $CSO(p,q,3-p-q)$ gauged maximal supergravity \cite{AlonsoAlberca:2003jq}. It can be viewed
as a contracted version of the $SO(p,8-p)$ gauged maximal supergravity in 4D, therefore having a
contracted gauge group and a different scalar potential. Note that this reduction provides an
example of gauge symmetry enhancement: the dimension of the $CSO(p,q,8-p-q)$ gauge group is larger
than three.

{From} the previous discussion it follows that the positive single exponential potentials of class
A, with $\Delta = 0,4$, can be embedded into gauged maximal supergravities in 4D with the following
gauge groups:
 \begin{align}
  \Delta = 0: \; CSO(1,1,6) \,, \qquad
  \Delta = 4: \; CSO(1,0,7) \,.
 \end{align}
We would also like to mention some results on other 4D gaugings, which are also of the $CSO$-form
but not obtainable by a 3D group manifold reduction. The following truncations to positive single
exponential potentials are possible \cite{Hull:1985rt, Gunaydin:1986cu, Townsend:2001ea}:
 \begin{align}
  \Delta = -3: \; CSO(4,4,0) \,, \, CSO(5,3,0) \,, \qquad
  \Delta = - 8/3: \; CSO(3,3,2) \,.
 \end{align}
The latter is the reduction of the $SO(3,3)$ gauged maximal supergravity in $D=5$. These values for
$\Delta$ correspond to pure cosmological constants, allowing for De Sitter solutions, in 4D and 5D
respectively. In \cite{Townsend:2001ea} it was already noted that the reduction of the 5D De Sitter
solution to 4D yields an accelerating cosmology corresponding to the critical point solution with
$\Delta = - 8/3$. The corresponding interpolating solution of section 2 will uplift to a 5D
solution which asymptotes to De Sitter.

Reduction of 7D maximal ungauged supergravity over a class B group manifold will also result in a
4D gauged maximal supergravity, which however lacks a Lagrangian. The same gauged supergravity can
be obtained by a $T^4$ reduction of the class B gauged supergravities in 8D
\cite{Bergshoeff:2003ri}. In analogy with table \ref{3Dalgebras} we consider these gauged
supergravities as deformations of class A with the deformation parameter $a$. Therefore, we will
denote their gauge groups by $CSO(p,q,8-p-q)_a$ with $p+q \leq 2$ (note the stricter range of $p+q$
as compared to class A; this follows from the Jacobi identity). With this understanding one can
assign the single exponential truncations of the class B results with $\Delta = -4/3,-1$ to the
following maximal gauged supergravities:
 \begin{align}
  \Delta = -\tfrac{4}{3}: \; CSO(0,0,8)_{a=1} ,\, CSO(2,0,6)_{a} \,, \qquad
  \Delta = -1: \; CSO(1,1,6)_{a=1/2} \,.
 \end{align}

\section{Comments} \label{comments}

We would like to comment on the relevance of the interpolating solutions to inflation. In this
context the number of $e$-foldings is crucial. It is defined by $N_e = \log (a (\tau_2) /
a(\tau_1))$ with $\tau_1$ and $\tau_2$ the start and end times of the accelerating period. These
times can easily be found in our approach as the points where the straight lines intersect the
``arctic circle". The number of $e$-foldings is required to be of the order of 65 to account for
astronomical data. For the interpolating solutions with $\Delta > -2$, which is a necessary
requirement to have a finite period of acceleration, one finds $N_e \lesssim 1$ \cite{Wohlfarth:2003ni,Ohta:2003ie,Chen:2003ij} for all values of
$a_1$, $a_2$ and $\Lambda$. The only exception to this behaviour is when $\Delta \rightarrow -2$,
where $N_e$ blows up. For the required 65 $e$-foldings one needs to take $\Delta + 2 \sim
10^{-60}$. As an example, for a compactification over an $m$-dimensional hyperbolic space, leading
to $\Delta = -2 + 2/m$, this translates into $m \sim 10^{60}$. Thus, it seems that the $e$-foldings
requirement for inflation can be met by a single exponential potential with a higher-dimensional
origin, be it admittedly tremendously much higher and evidently not in the context of M- or string
theory. For the present-day acceleration, however, solutions for exponential potentials from
M-theory could be relevant. \\

In section 3 we have discussed reduction over the Bianchi group manifolds. There we have taken the
universal cover of the group manifold. For this reason, types I--VIII are non-compact and have the
topology of $\mathbb{R}^3$ while type IX has the topology of $S^3$. Therefore, the latter case does
not raise any issues when compactifying. In the case of non-compact group manifolds there are two
approaches: (1) Supersymmetry is preserved but one can only use the reduction to uplift solutions
since the non-compact internal manifold leads to a continuous spectrum in the 4D theory, which is
not physically acceptable. This is the so-called non-compactification scheme; (2) the group
manifold is compactified by dividing out by discrete symmetries \cite{Thurston:1979}. For all
Bianchi types except types IV and VI$_a$ it is possible to construct compact manifolds in this way
\cite{Barrow:2000ka}. Sometimes supersymmetry is preserved under this operation, like for the
three-torus, but sometimes it is not, like for a hyperbolic manifold. In the latter case one
clearly cannot embed the system (\ref{Lagrangian2}) in a gauged supergravity. \\

Our analysis of the triple exponential potentials suggests the existence of exotic $S(D-3)$-branes
in $D$ dimensions. The situation with one standard and two exotic solutions stemming from the
different twisted $SL(2,\mathbb{R})$ reductions is analogous to the structure of domain wall
solutions in these systems. This was analysed in the case of the reduction of IIB supergravity to
9D gauged maximal supergravity. The three distinct domain walls uplift to one standard and two
exotic half-supersymmetric 7-brane solutions of IIB \cite{Bergshoeff:2002mb}. These carry charges
that take values in one of the three inequivalent subgroups of $SL(2,\mathbb{R})$. The standard
case is the D7-brane and corresponds to a charge vector in the $\mathbb{R}$ subgroup of
$SL(2,\mathbb{R})$. The exotic cases have charge vectors that take values in the other two
subgroups of $SL(2,\mathbb{R})$ and can be interpreted as a D7-Q7 or a D7-$\overline{\text{Q7}}$
bound state \cite{Bergshoeff:2002mb}, where the Q7 is the S-dual of the D7.

A similar situation arises with instantons where we also find solutions with charge vectors in all
three subgroups of $SL(2,\mathbb{R})$ \cite{Bergshoeff:2003ab}. Note that these solve the same
axion-dilaton-gravity system as the S-branes and the 7-branes, with the only difference being that
in the case of the instantons we work with a Euclidean version. Again, the standard D-instanton has
a charge vector that takes values in the $\mathbb{R}$ subgroup of $SL(2,\mathbb{R})$. The
interpretation of the exotic instanton-like solutions corresponding to the other two subgroups of
$SL(2,\mathbb{R})$ is less clear \cite{Bergshoeff:2003ab}.

In the case of the S-branes it would be interesting to see whether, analogously to the 7-branes and
instantons, solutions corresponding to each of the three subgroups can be constructed. This would
mean the existence of new ``exotic'' S-branes, which have not been given in the literature. \\

Finally, it would be interesting to extend our work to more general potentials with multiple
scalars. It has been argued that multi-exponential potentials with orthogonal dilaton couplings
lead to assisted inflation for the power-law solutions \cite{Liddle:1998jc}. It would be very
interesting to investigate the consequences for the interpolating solutions as well. Other
simplifications might occur for the case when the dilaton couplings form the Cartan matrix of a
semi-simple Lie group; in such cases the system becomes an (integrable) Toda model
\cite{Gavrilov:1994sv, Lukas:1997iq, Lu:1997jr}.

\section*{Acknowledgements}

We thank Marc Henneaux and Rom\'{a}n Linares for interesting discussions. This work is supported in
part by the European Community's Human Potential Programme under contract HPRN-CT-2000-00131
Quantum Spacetime, in which E.B., A.C., M.N.~and D.R.~are associated to Utrecht University. The
work of U.G. is funded by the Swedish Research Council.

\bibliography{accelerating}
\bibliographystyle{utphysmodb}

\end{document}